\documentclass{aa}  

\usepackage{graphicx}
\usepackage{txfonts}
\usepackage{xcolor}
\usepackage{footmisc}

\newcommand{\micron}{\ensuremath{\mu m}}
\newcommand{\Teff}{$\mathrm{T}_\mathrm{eff}$}
\newcommand*\chem[1]{\ensuremath{\mathrm{#1}}}
\newcommand{\matr}[1]{$\mathbf{#1}$}

\graphicspath{{./Figures}{}}
\RequirePackage[normalem]{ulem} 
\RequirePackage{color}\definecolor{RED}{rgb}{1,0,0}\definecolor{BLUE}{rgb}{0,0,1}\definecolor{BLACK}{rgb}{1,1,1}

\begin{document}

   \title{
   Spectral unmixing for exoplanet direct detection in hyperspectral data}

   \author{J. Rameau
          \inst{1}
          \and
          J. Chanussot\inst{2}
          \and
          A. Carlotti
          \inst{1}
          \and
          M. Bonnefoy
          \inst{1}
          \and
          P. Delorme
          \inst{1}
          }

   \institute{Univ. Grenoble Alpes, CNRS, IPAG, F-38000 Grenoble, France\\
              \email{julien.rameau@univ-grenoble-alpes.fr}
         \and
             Univ. Grenoble Alpes, Inria, CNRS, Grenoble INP, GIPSA-Lab, Grenoble, F-38000, France
             }

   \date{Received January 13th, 2021; Accepted March 18th, 2021}

 \abstract{The direct detection of faint exoplanets with high-contrast instruments can be boosted by combining it with high spectral resolution. For integral field spectrographs yielding hyperspectral data, this means that the majority of the field of view consists of diffracted starlight spectra and a spatially localized planet. Observation analysis usually relies on classic cross-correlation with theoretical spectra, maximized at the position and with the properties of the planet. In a purely blind-search context, this supervised strategy can be biased with model mismatch and/or be computationally inefficient.}{Using an approach that is inspired by the analysis of hyperspectral data within the remote-sensing community, we aim to propose an alternative to cross-correlation that is fully data-driven, which decomposes the data into a set of individual spectra and their corresponding spatial distributions. This strategy is called spectral unmixing.}{We used an orthogonal subspace projection to identify the most distinct spectra in the field of view. Their spatial distribution maps were then obtained by inverting the data. These spectra were then used to break the original hyperspectral images into their corresponding spatial distribution maps via non-negative least squares. A matched filter with the instrument point-spread function (or visual inspection) was then used to detect the planet on one of the maps. The performance of our method was evaluated and compared with a cross-correlation using simulated hyperspectral data with medium resolution from the ELT/HARMONI integral field spectrograph.}{We show that spectral unmixing effectively leads to a planet detection solely based on spectral dissimilarities at significantly reduced computational cost. The extracted spectrum holds significant signatures of the planet while being not perfectly separated from residual starlight. The sensitivity of the supervised cross-correlation is three to four times higher than with unsupervised spectral unmixing, the gap is biased toward the former because the injected and correlated spectrum match perfectly. The algorithm was furthermore vetted on real data obtained with VLT/SINFONI of the $\beta$ Pictoris system. This led to the detection of $\beta$ Pictoris b with a signal-to-noise ratio of 28.5.}{Spectral unmixing is a viable alternative strategy to a cross-correlation to search for and characterize exoplanets in hyperspectral data in a purely data-driven approach. The advent of large data from the forthcoming IFS on board \textit{JWST} and future ELTs motivates further algorithm development along this path.}


   \maketitle
%
%-------------------------------------------------------------------

\section{Introduction}

Direct detection of extrasolar planets relies on a combination of high angular resolution to spatially resolve the point-like planet to its host star at the subarcsecond level, high contrast to suppress the starlight, which is several orders of magnitudes brighter than the exoplanet, and optimized wavefront control to attenuate the diffracted starlight. The sensitivity to faint and/or close-in exoplanets is nevertheless plagued by residual starlight stemming from uncalibrated instrumental aberrations, and for ground-based facilities, Earth's atmospheric turbulence. Disentangling the exoplanet from these so-called speckles is commonly achieved with differential observing techniques. Most of these strategies introduce relative spatial motion, azimutal (ADI) and/or radial (SDI), between an astrophysical source and the speckle field \citep{Racine:1999, Marois:2003}. This spatial diversity is the central concept of post-processing algorithms of different complexities that model and subtract the speckles while preserving the signal of any planet \citep{Marois:2006, Lafreniere:2007, Soummer:2012, Amara:2012, Gomez:2016,Ren:2018,Flasseur:2018,Samland:2020}.

The ground-based extreme adaptive-optics systems Gemini/GPI \citep{Macintosh:2014}, the Very Large Telescopes VLT/SPHERE \citep{Beuzit:2019}, Subaru/SCExAO-CHARIS \citep{Jovanovic:2015}, and the LBTI with ALES \citep{Esposito:2011,Skemer:2015} make best use of the high performance of these key aspects to detect exoplanets and infer their individual and population properties \citep[e.g.,][]{Macintosh:2015,Chauvin:2017,Keppler:2018,Nielsen:2019, Vigan:2020}. They are in particular equipped with an integral field spectrograph (IFS), providing two spatial and one spectral dimensional images, or hyperspectral data. The spectral range ($\Delta\lambda\sim0.3-1.5~\micron$) is large enough to benefit from the radial motion of the speckle field with the diffraction to improve its subtraction and hence the sensitivity to faint exoplanets \citep{Mesa:2015,WangJ:2015,Ruffio:2017,Brandt:2017,Galicher:2018}. The IFS further plays a key role in studying the atmosphere of the planets by providing a spectrum with low spectral resolution ($R=\lambda/\delta\lambda\sim20-75$)  of each detection. At this resolution, broad molecular absorption bands can be identified, and the pseudo-continuum holds information about the pressure-temperature profile and clouds \citep[e.g.,][]{DeRosa:2016,Bonnefoy:2016,Chilcote:2017,Rajan:2017,Delorme:2017,Samland:2017,Bonnefoy:2018,Greenbaum:2018,Muller:2018,Chauvin:2018,Currie:2018,Uyama:2020,Stone:2020, Ward:2021}.

A higher spectral resolution ($R>5000-100,000$) is a way to proceed in our understanding of the physical, chemical, and orbital properties of imaged exoplanets. Resolving the absorption lines augments the information content of the spectra and provides access to radial velocity \citep{Snellen:2014,Wang:2018,Ruffio:2019}, spin \citep{Snellen:2014,Schwartz:2016,Bryan:2018}, element abundances \citep{Konopacky:2013,Barman:2015,Wilcomb:2020,Petrus:2020}, or accretion tracers \citep{Haffert:2019}. The combination of high spectral resolution and high contrast can also significantly enhance the sensitivity to faint exoplanets \citep{Sparks:2002,Riaud:2007,Kawahara:2014,Snellen:2015,Lovis:2017,Wang:2017}. The technique assumes that the planet spectrum and that of the residual starlight can be separated through distinct absorption lines and/or different Doppler shifts that are due to orbital motion. The detection algorithm relies on straightforward cross-correlations of the signal with model spectra of the planet, which sum all common lines and cancel out nonplanetary features. The method has been proven to be very efficient even on instruments that were not designed to reach high contrast: the slit spectrographs VLT/CRIRES and Keck/NIRSpec \citep{Snellen:2014,Wang:2018}, and the medium-resolution IFS Keck/OSIRIS and VLT/Spectrograph for INtegral Field Observations in the Near-Infrared SINFONI \citep{Barman:2011,Konopacky:2013,Barman:2015,Hoeijmakers:2018,PetitdelaRoche:2018,Ruffio:2019,Wilcomb:2020,Petrus:2020}. This strategy fuels the development of a new generation of instruments on existing 8- to 10-meter-class telescopes \citep{Kuntscner:2014,Mawet:2018,Otten:2021} and on future extremely large telescopes (ELTs) \citep{McGregor:2012,Thatte:2016,Brandl:2018}. 

As of today, the cross-correlation technique has been applied on previously identified exoplanets. It was employed to extract new measurements of different planet characteristics that are accessible at medium to high spectral resolution. It has not been used for pure detection purposes. The prior knowledge of the planet that is searched for is a serious advantage in narrowing down the grid search of model spectra that are to be cross-correlated with the signal. Without such priors, the library should be large enough to encompass all possible spectral signatures, in particular to explore a wide range of effective temperatures and surface gravities that drive the emitted spectrum of a given molecule as well as radial velocities that control the positions of the lines. Still, the method relies on theoretical models that must resemble true planet spectra \citep[e.g.,][]{Petrus:2020}. In this sense, it falls into the regime of \textit{\textup{supervised}} source-detection algorithms. It can also quickly become computationally time consuming for large libraries and/or large detectors. For these two reasons, unsupervised and fast alternative algorithms should be investigated to provide an alternative to the cross-correlation technique. The advent of large hyperspectral data with the IFS on board the \textit{JWST} \citep{Rieke:2015,Bagnasco:2007} and even more with ELTs-IFS such as the High Angular Resolution Monolithic Optical and Near-Infrared Integral field spectrograph HARMONI \citep{Thatte:2016}, METIS \citep{Brandl:2018}, GMTIFS \citep{McGregor:2012}, and IRIS \citep{Larkin:2016}, motivates this development.

Blind source-detection in hyperspectral data, or spectral unmixing for this type of data, represents a long-standing problem beyond astronomy, notably in Earth remote-sensing and microscopy. The dedicated algorithms aim at separating various sources in the field of view from their distinct spectral morphologies in a purely data-driven approach. Spectral umixing is a very active research field to develop powerful algorithms with growing complexities and specific features for a given science case, especially for Earth observations \citep[see a review of the main methods from][and references therein]{Bioucas-Dias:2012}. In astronomy, spectral unmixing has been used to analyze IFS data from \textit{Spitzer, Herschel}, or planetary probes. Some applications include velocity maps of interstellar clouds \citep{Juvela:1996}, interstellar dust spectral inference \citep{Rapacioli:2005,Berne:2007,MOUSSAOUI20082194,Gratier:2017,Foschino:2019,Boulais:2020}, or Mars surface mapping \citep{Forni:2005,Hauksdottir06thephysical,themelis:2011,Liu:2018}. Beyond these examples, methods like this are rare in astronomy to our knowledge despite the amount of existing and forthcoming hyperspectral data, and they have never been tested for exoplanet science.

In this paper, we explore the use of unsupervised spectral umixing as an alternative to the supervised cross-correlation technique to directly detect an exoplanet and extract its spectrum from medium-resolution data. We test one algorithm on synthetic $R\sim7000$ K-band HARMONI data to evaluate its performances. We then apply it on on-sky $R\sim5000$ K-band SINFONI data of $\beta$ Pictoris and show that spectral umixing can effectively detect $\beta$ Pictoris b in less than a minute on a regular laptop. The paper is organized as follows. In section \ref{sec:maths} we motivate the unmixing approach for direct exoplanet detection in hyperspectral data and provide the mathematical framework of the adopted algorithm. In section \ref{sec:harmoni} we describe the simulations of the HARMONI data and demonstrate the algorithm on a simple test case, after which detection sensitivity and spectral fidelity are discussed and compared to the cross-correlation technique in section \ref{sec:perf}. In section \ref{sec:sinfoni} the algorithm is applied on the SINFONI data. Advantages and limitations of the algorithm are discussed in Sect. \ref{sec:discussion}, followed with concluding remarks and proposals for improvement.

%--------------------------------------------------------------------
\section{Source separation from spectral decomposition}
\label{sec:maths}

\subsection{Statement of the problem}
\begin{figure}
    \centering
    \includegraphics[width=0.5\textwidth]{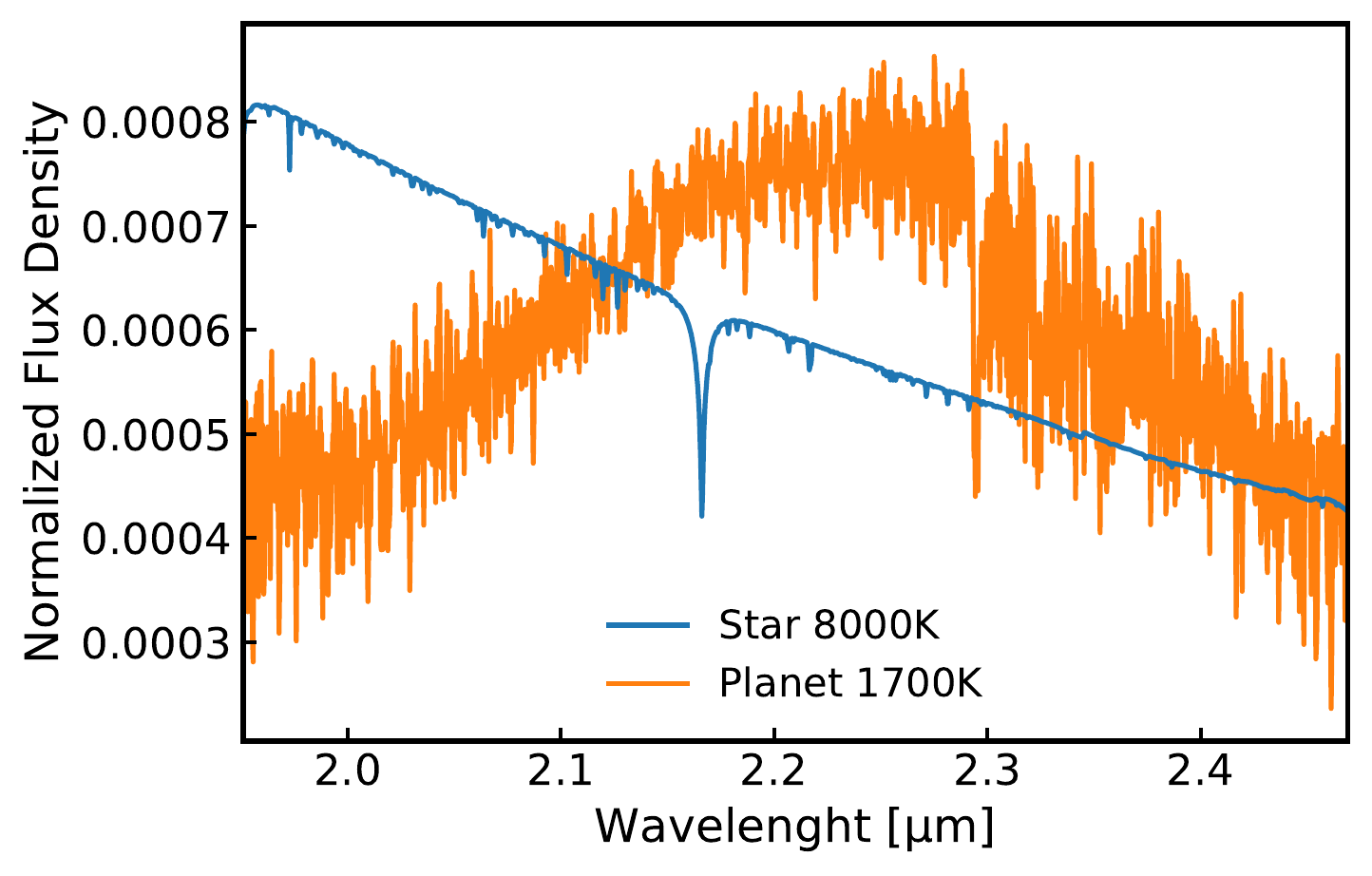}
    \caption{Model spectra of a $\beta$ Pictoris-like system: An A6 star with \Teff$=8000$\,K \citep[BT-Nextgen][]{Allard:2012} and a giant planet with \Teff$=1700$\,K \citep[BT-Settl][]{Allard:2014}, normalized over the K bandpass at a resolution of $R=7000$. Not only the slope of the continuum is reversed between the two objects, but also the absorption line content, with the prominent Brackett-$\gamma$ line in the stellar spectrum at $2.16$\,\micron,~while the planet spectrum exhibits a forest of \chem{H_2O} and \chem{CO} lines.}
    \label{fig:model_spec}
\end{figure}

A typical hyperspectral exoplanet imaging dataset contains two sources: the star and the orbiting planet. The two astrophysical objects have different \Teff, from 2800\,K up to $>$10000\,K for the former, and $<$2000\,K for the latter. The temperature drives the formation of elements in their atmospheres (hydrogen- and metal-bearing ions, atoms, and molecules for stars, and \chem{H_2O}, carbon- and nitrogen-bearing molecules for planets), which hence plays a key role in the appearance of the emitted spectra. As illustrated in Figure \ref{fig:model_spec} in the K band, the two spectra are distinct at low frequency \footnote{"Frequency" is used here and throughout to describe the shape of spectra at different spectral scalings, and it should not be confused with the inverse of the unit wavelength.} by the blackbody emission and the broad absorption bands and at high frequency by the various absorption lines. This dissimilarity can be exploited to separate the two sources in the hyperspectral data. 

While the planet is spatially localized in the image, the stellar spectrum is spatially replicated over the image via the diffracted residual starlight. A given spaxel \matr{d_i} in the hyperspectral data can be modeled with a linear statement, such that
\begin{equation}
\mathbf{d_i} = \sum_{j=1}^{p}\mathbf{s_j}\alpha_{j,i}+\mathbf{\epsilon_i}\,,
\label{eq:lm_pix}
\end{equation}
where \matr{s_j} is the spectrum of the $j$th source, $\alpha_{j,i}$ is the corresponding weight at the $i$th spatial location, and \matr{\epsilon_i} is an additive noise term. When we assume that the source spectra are spatially invariant in the image, that is, all pixels can be expressed a linear combination of the same $p$ spectra, equation \ref{eq:lm_pix} can be rewritten in compact matrix form as
\begin{equation}
    \mathbf{D}=\mathbf{S}\mathbf{A}+\mathbf{E}\,,
\label{eq:lmm}
\end{equation}
where \matr{D} $\in\mathbb{R}^{n_\lambda\times m}$ is the column-vectorized hyperspectral images where the columns contain the $m=n_x \times n_y$ spaxels in any order of $n_\lambda$ spectral channels, \matr{S} $\in\mathbb{R}^{n_\lambda\times p}$ is a matrix of $p$ source spectra, \matr{A} $\in\mathbb{R}_+^{p\times m}$ is the matrix of their corresponding spatial weight maps, and \matr{E} the uncorrelated noise. Equation \ref{eq:lmm} is the typical linear mixture model assumed for spectral unmixing. \matr{S} contains the so-called endmembers, and \matr{A} contains the abundance maps \citep{Bioucas-Dias:2012}.

The decomposition requires three steps: i/ the estimation of the dimensions of \matr{S} and \matr{A}, that is, the number $p$ of spectra in the data, ii/ the extraction of the spectra, and iii/ the evaluation of the weight maps. Each step is detailed in the following.

\subsection{Estimation of the number of sample spectra}

\begin{figure}
    \centering
    \includegraphics[width=0.5\textwidth]{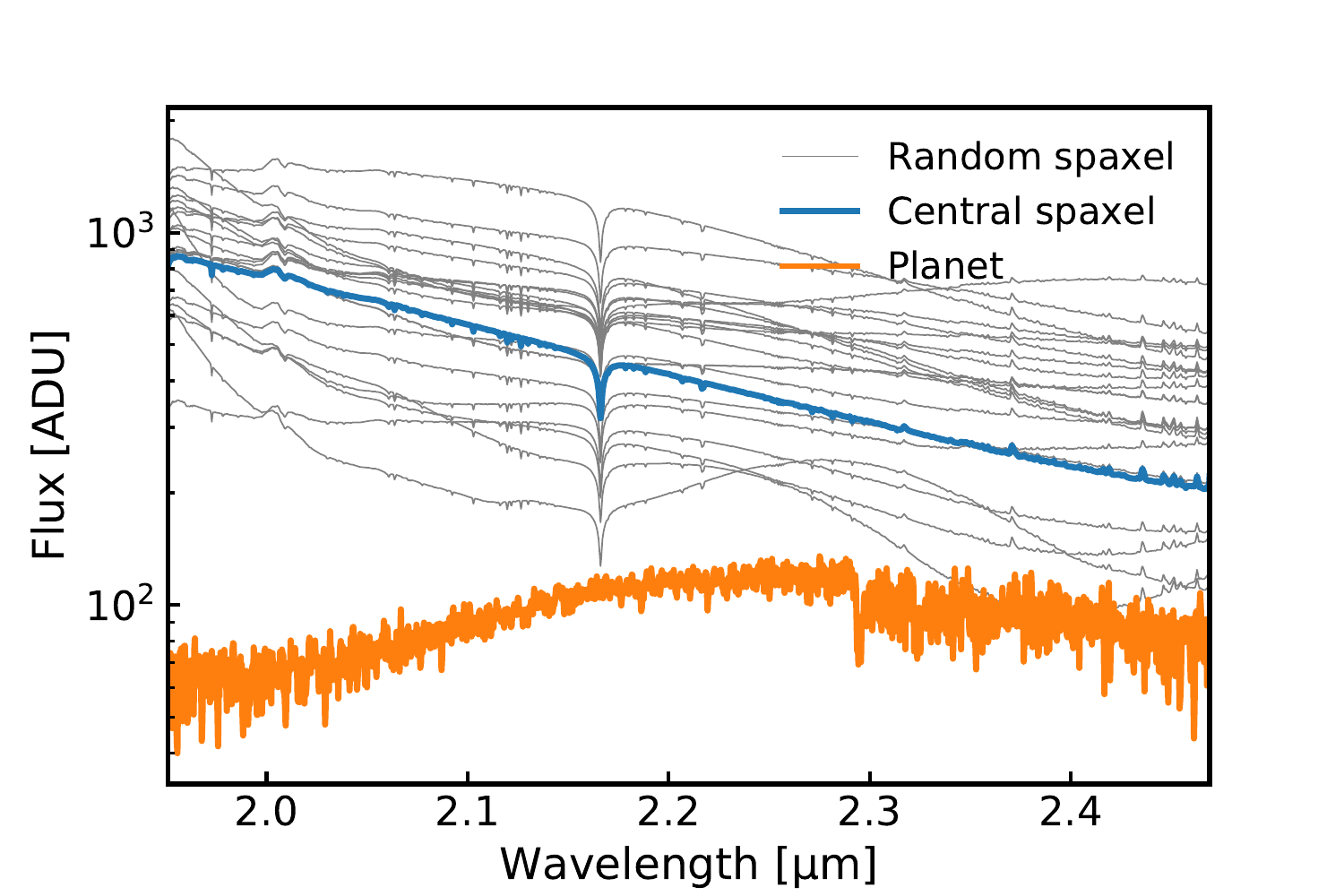}
    \includegraphics[width=0.5\textwidth]{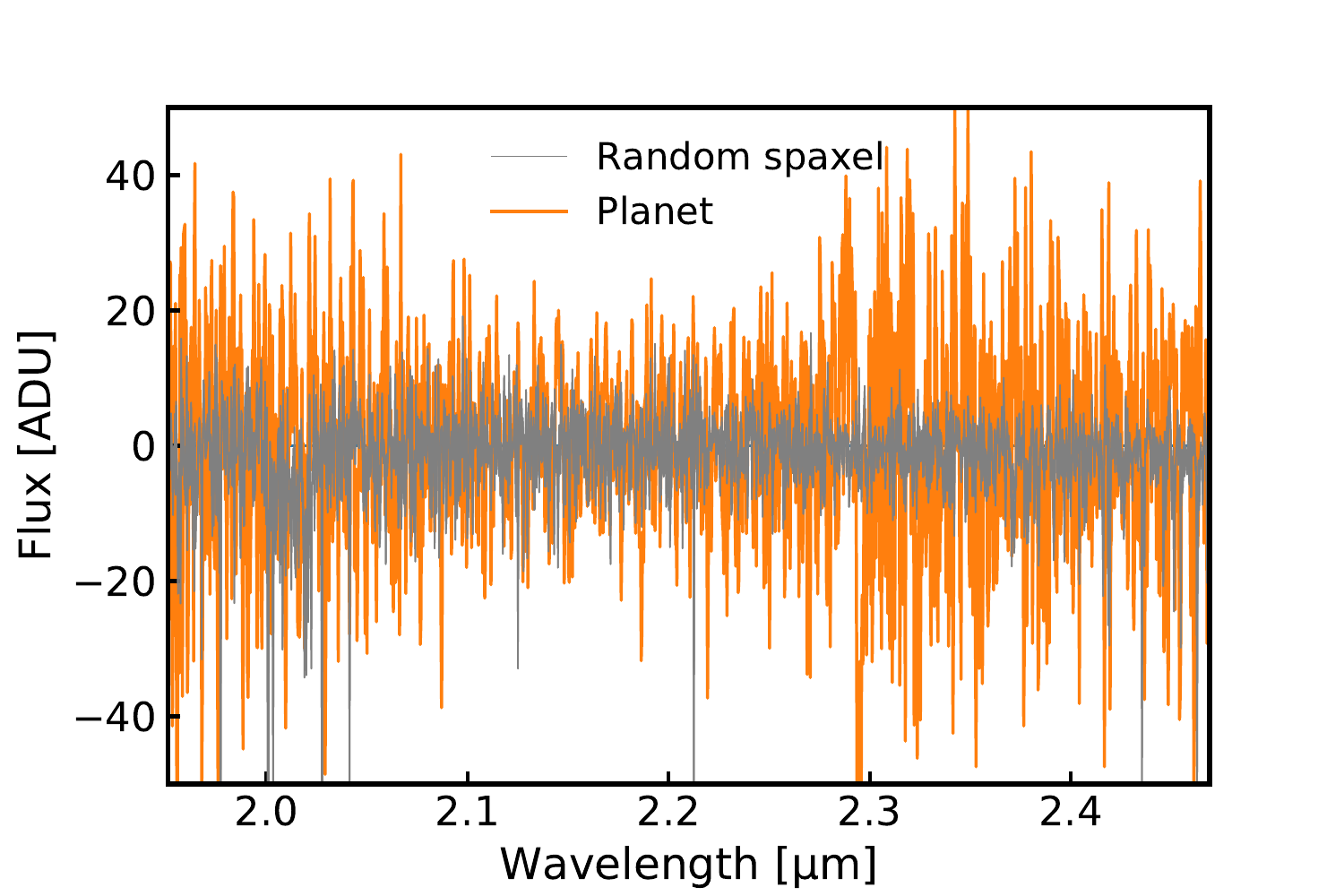}
    \caption{[Top:] Spectra of individual spaxels in synthetic coronagraphic HARMONI data at $R\sim7000$ with a central A6 star and a $1700$\,K planet (orange) (see Figure \ref{fig:model_spec}). Random spaxels (gray) in the corrected region illustrate the modulation by the speckles of the stellar spectrum, as shown with the central spaxel behind the semitransparent focal plane mask (blue). Details of the simulations are given in section \ref{sec:harmoni}. [Bottom:] Resulting spectra after the continuum was removed (following \citep{Hoeijmakers:2018}). The amplitude of the planet spectra has been amplified for visualization purposes.}
    \label{fig:harmoni_spec}
\end{figure}
The initial hypothesis of only two sources, the star and the planet, does not hold in high-contrast data. The speckles act as modulators of the stellar spectrum as a function of position and wavelength, as illustrated in Figure \ref{fig:harmoni_spec} for synthetic HARMONI data (see section \ref{sec:harmoni} for details on the simulation). Therefore the \textit{\textup{effective}} number of samples in our hyperspectral data can be as high as the number of speckles in the field of view. This number might further be augmented by the spatially variable imperfect calibration of the data (wavelength, crosstalk, and bad pixels) or line spread function, as in VLT/MUSE data \citep{Xie:2020}\footnote{Because of these variations, nonplanetary sample spectra should be interpreted with caution and are not to be considered as perfect stellar spectra.}.

Estimating the number of sources in hyperspectral data is a difficult task in spectral unmixing and subject to regular algorithm development. Several techniques have been proposed to exploit the fact that the inherent structure of high-dimension hyperspectral data usually lies in a lower-dimensional subspace \citep[e.g.,][]{Chang:2006,Bioucas-Dias:2008,Acito:2009,Luo:2013,Deville:2014,Halimi:2015,Drumetz:2016}. These methods were designed for the low-contrast regime, however, and are sensitive to outliers, noise, and spectral variation across the field of view (referred to as variability) \citep[][and references therein]{Drumetz:2020}. 

Because of these limitations and because only the extraction and identification of the planet component matters, the maximum number of \textit{\textup{effective}} samples $p$ is at the moment set empirically by the user. Because the algorithm has a low time complexity (see below), we recommend running a first trial with a large p of $15-20$ or more in case of large images (making sure $p\ll n_\lambda$, and $p\ll m=n_x*n_y$, for the algorithm to work). p can then be adjusted to the minimum number of samples necessary to detect any planet in the data. For the data set used in this work, which was tested with a range of contrasts and separations, all planets were extracted from the first 2-10 components. Therefore $p=10$ was chosen for synthetic HARMONI and real SINFONI data in the following; the decomposition into a larger number of samples results in additional starlike spectra, which is not necessary for planet detection. Because the algorithm is very fast (see below), it is not necessary to explore complex algorithms suitable for estimating $p$ for our data. We therefore defer this to a future work. 

\subsection{Orthogonal subspace projection to extract the sample spectra}

The extraction of the sample spectra is a very active topic in remote-sensing research; many strategies have been investigated based on geometrical or statistical considerations \citep[see the reviews by ][and references therein]{Plaza:2006,Bioucas-Dias:2015}. We considered a geometry-based algorithm, which is the most frequently used algorithm family for spectral unmixing applications. This approach finds the set of the purest spaxels in the data for each requested sample. 

Except for planets that are as bright or brighter than the stellar halo, this assumption does not hold at high contrast. We circumvented the problem by subtracting the halo in the spectral dimension with data preprocessing. As discussed before, the halo mostly affects the broadband flux and the continuum of the spaxels. It is therefore a low spectral and spatial frequency component that can be removed with several strategies such as high-pass filtering \citep{Konopacky:2013,Ruffio:2019}, continuum normalization \citep{Hoeijmakers:2018,Haffert:2019,Petrus:2020}, or principal component analysis \citep{Xie:2020}. Removing the low frequencies of the stellar halo in the spectral dimension also minimizes the variation in the stellar spectrum across the field of view, which in turn reduces the effective number of samples. As a consequence, this procedure also removes the continuum of the planet spectrum (see the example following \citealt{Hoeijmakers:2018} in Figure \ref{fig:harmoni_spec}). However, the distinct and multiple absorption lines between the different spectra enables spectral unmixing in the high-frequency components.

Following equation \ref{eq:lmm} and with $p\ll n_\lambda$, all spaxels live in a low-dimensional subspace, or simplex, whose vertices (geometrical corners)  are represented by the sample spectra. A given vertex is therefore most distant to the subspace spanned by the other $p-1$ vertices. An orthogonal subspace projection (OSP) approach can be used to identify the vertices \citep[e.g.,][]{Harsanyi:1994, Chang:2005}. 

Let \matr{s_0} be the first sample spectrum. The sparsity of the planet over the diffracted starlight in the field of view motivates the choice of \matr{s_0} as the spectrum of the median spaxel. Let \matr{U} be the spectral sample matrix, which at this step only contains \matr{s_0}, such that \matr{U}$=[\mathbf{s_0}]$. The orthogonal projector from \matr{U}, denoted $P_\mathbf{U}^\perp$, is defined as
\begin{equation}
\label{eq:projector}
    P_\mathbf{U}^\perp = \mathbf{I}-\mathbf{U}(\mathbf{U}^T\mathbf{U})^{-1}\mathbf{U}^T\,,
\end{equation}
with \matr{I} a $n_\lambda\times n_\lambda$ identity matrix, $\mathbf{U}^T$ the transpose of \matr{U}, and $(\mathbf{U}^T\mathbf{U})^{-1}\mathbf{U}^T$ the pseudo-inverse of \matr{U}. The dot product $P_\mathbf{U}^\bot\mathbf{d_i}$ projects each spaxel \matr{d_i} in $\langle\mathbf{s_0}\rangle^\bot$, the subspace orthogonal to the space linearly spanned by \matr{s_0}\footnote{$\langle\mathbf{s_0}\rangle^\bot$ is defined such that $\forall\mathbf{v}\in \langle\mathbf{s_0}\rangle^\bot$, $\mathbf{v}\cdot\mathbf{s}_0$=0.}. Iterating over all spaxels, a second sample spectrum, set as \matr{s_1}, is flagged as the most distant spaxel to \matr{s_0}. The projection distance is chosen to be the sum of the squared residuals after removing from \matr{d_i} the projection of \matr{d_i} onto \matr{s_0}, or 
\begin{equation}
\label{eq:distance}
(P_\mathbf{U}^\perp\mathbf{d_i})^T(P_\mathbf{U}^\perp\mathbf{d_i})\,.
\end{equation}
A new orthogonal subspace projector $P_\mathbf{U}^\perp$ can be defined from the linear subspace spanned with \matr{U}$=[\mathbf{s_0},\mathbf{s_1}]$ and applied to the remaining spaxels. The third sample spectrum is identified as the one with the maximum absolute orthogonal projection in the subspace $\langle\mathbf{s_0,s_1}\rangle^\perp$\footnote{The sample spectra do not form an orthogonal basis, $^\perp$ denotes the orthogonality of the space to the origin vectors.}. The process is repeated until $p$ sample spectra are exhausted to build up \matr{S}. To prevent bad or hot pixels from being flagged as sample spectra because they differ by construction from the remaining spaxels, they must be properly corrected for during data preprocessing.

\subsection{Evaluation of the weight maps}

From the extraction of the matrix \matr{S} of sample spectra, the determination of the spatial weight matrix \matr{A} from equation \ref{eq:lmm} is an inverse problem, which can be solved with least-squares means. The problem can be further constrained because all spaxels in our hyperspectral data are a positive linear combination of the sample spectra, that is, \matr{A}$\geq0$. Therefore the non-negative least-squares estimate of \matr{A} is done for each spaxel \matr{d_i} by finding
\begin{equation*}
    \mathrm{arg}\min_\mathbf{\alpha_i}\vert\vert \mathbf{S\alpha_i}-\mathbf{d_i}\vert\vert_2\,, \mathrm{subject}\medspace\mathrm{to}\medspace\mathbf{\alpha_i}\geq0\,,
\end{equation*}
where $\vert\vert\cdot\vert\vert_2$ denotes the Euclidian norm. 

We used the \texttt{optimize.nnls} module from \texttt{scipy} to solve the problem. In practice, this iterative algorithm is very slow to converge. Because in our case the number of coefficients is much smaller than the number of wavelength channels, that is, $p\ll n_\lambda$, the solution proposed by \citet[(proofs therein)]{Bro:1997} can be used such that solving the non-negative least-squares problem is accelerated by replacing \matr{S} with a $p \times p$ matrix following
\begin{equation}
    \label{eq:nnls}
    \mathrm{arg}\min_\mathbf{\alpha_i}\vert\vert (\mathbf{S}^T\mathbf{S})\mathbf{\alpha_i}-\mathbf{S}^T\mathbf{d_i}\vert\vert_2\,, \mathrm{subject}\medspace\mathrm{to}\medspace\mathbf{\alpha_i}\geq0\,.
 \end{equation}
 Repeating the resolution over the entire image builds \matr{A}. Reshaping the matrix into the original 2D spatial dimensions finally generates $p$ weight maps corresponding to the decomposition of the input hyperspectral image \matr{D}. 

\subsection{Planet detection and summary of the algorithm }
The weight maps provide the spatial distribution of each sample in the data in an undetermined way; the map associated with the planet component has to be identified a posteriori. The spatial signature of the planet is defined by the point-spread function (PSF) with a finite spatial extent (the full width at half maximum, FWHM) that is typically larger than a single pixel. Planet vetting can be performed visually because the number of maps is usually small, or it can be conducted automatically using a matched filter \citep{Cantalloube:2015,Ruffio:2017}.

The algorithm based on spectral unmixing for detecting a planet in hyperspectral data is summarized below.
\begin{enumerate}
    \item Set $p$ as the number of \textit{\textup{effective}} sample  spectra in the data ($p>2$ due to spatial variability). $p=15-20$ might be a good first guess to prevent missing any planet, while a higher value is recommended for very large images. For the data explored in this work, $p=8-10$ is appropriate.
    \item Remove the continuum for every spaxel to keep the high-frequency components.
    \item Vectorize the hyperspectral data \matr{D} in a 1D spatial plus 1D spectral matrix. 
    \item Set the first sample \matr{s_0} as the closest spectrum to the median spaxel and define \matr{U}=[\matr{s_0}] the subspace spanned by \matr{s_0}.
    \item Project each spaxel onto the linear subspace spanned by \matr{U} with the orthogonal projector $P_\mathbf{U}^\perp$ defined in equation \ref{eq:projector}.
    \item Find the spaxel with the extremal projection (see equation \ref{eq:distance}), which corresponds to a new sample spectrum, \matr{s_\mathrm{new}}.
    \item Append \matr{U} with \matr{s_\mathrm{new}}.
    \item Repeat steps 5 to 7 until all $p$ spectra are extracted.
    \item Invert \matr{D} with fast non-negative constrained least squares (see equation \ref{eq:nnls}) to compute the spatial distribution of each sample spectrum \matr{A}.
    \item Expand \matr{A} in $p$ 2D spatial maps to search for a planet.

\end{enumerate}

\section{Application on synthetic data}
\label{sec:harmoni}

In this section, the capability of the proposed unsupervised spectral unmixing algorithm is first tested on synthetic high-contrast hyperspectral HARMONI data. Then we compare it with classic cross-correlation.

\subsection{Simulations of ELT/HARMONI data}

HARMONI will be the first-light IFS on the ELT offering diffraction-limited hyperspectral images at medium spectral resolution ($3500-18000$) in the optical and near-infrared \citep{Thatte:2016}. A module consisting of an apodized pupil coronagraph and a ZELDA wavefront sensor will supplement the single conjugate adaptive optics of the core instrument and give access to high contrast ($>10^{-5}$) at very short separations ($\ge$ 50 mas) in the Nyquist-sampled $0.61\,"\times0.86\,"$ field of view with a spatial scale of $4$ mas/pixel \citep{Carlotti:2018}. 

A synthetic coronagraphic sequence is computed with a dedicated end-to-end simulator with the following configuration: 4h of observations with one-minute-long exposures in angular differential imaging for a star at a declination of $-15$\,deg under median seeing condition ($0.7-0.8$\,") at K band ($1.951-2.469\,\micron$, $R=7104$) with the shaped pupil SP1 (working angles in the $5-20\,\lambda/D$ range) and the partly opaque focal plane mask FPM1. This enables direct calibration with the attenuated ($10^{-4}$) central star. All details of the simulator are presented in \citet{Carlotti:2018} in Appendix \ref{ap:sim}. 

Planets can further be injected into the mock data cubes. The off-axis simulated PSF is used as a template for a simulated planet, and a BT-Settl model spectrum \citep{Allard:2014} with an effective temperature of $1700$\,K and $\log g=4.0$\,dex is assumed by default in the following, unless specified. These parameters are typical of known L-type young giant exoplanets (HIP\,65426\,b, $\beta$ Pictoris\,b). For simplicity, no Doppler shift is assumed. The planet has the same radial velocity as the central star. Contrast and atmospheric transmission are calibrated with the attenuated central star. These settings are kept constant for all the following simulations, only contrast and injected position vary. To speed up the injection and recovery processes in the following, a 2h subset of the simulated data was used and the central $100\times100$ pixels were cropped in each frame, resulting in $100\times100\times1665\times120$ cubes for a total of 16 Gb.

\subsection{Single test case}
\begin{figure*}[th]
\centering
 \includegraphics[width=0.35\textwidth]{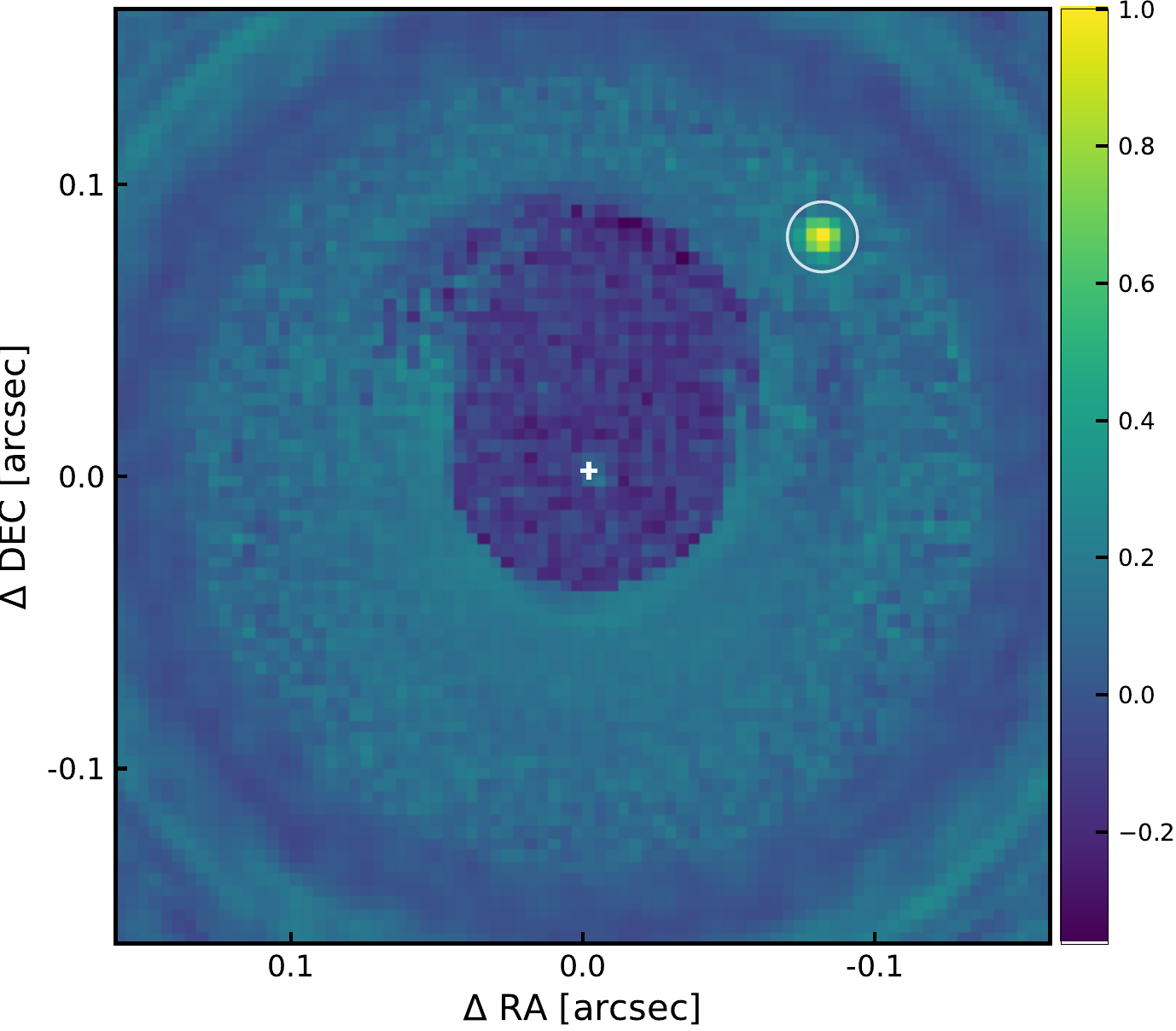}
   {\includegraphics[width=0.35\textwidth]{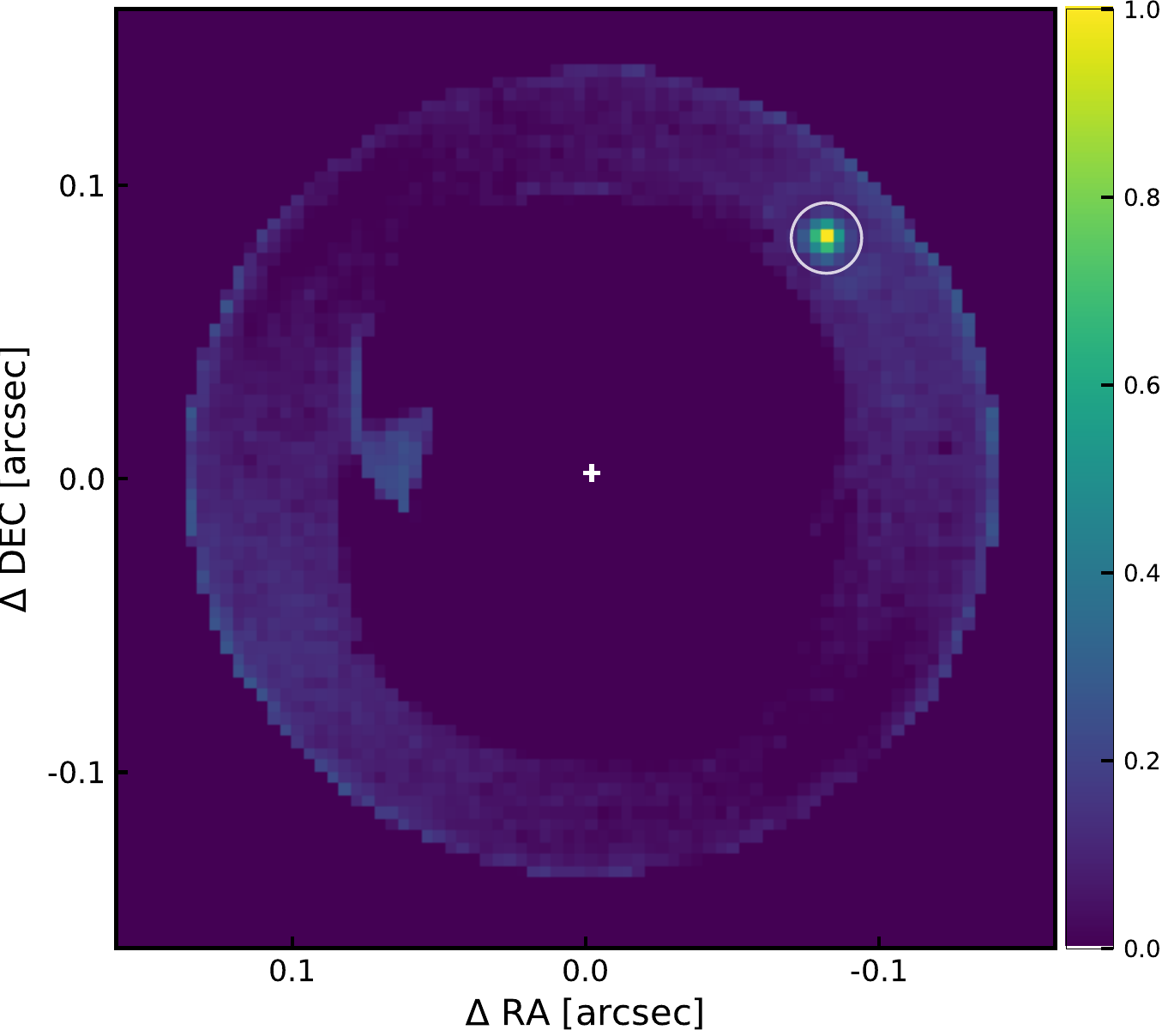}}
    {\includegraphics[width=0.25\textwidth]{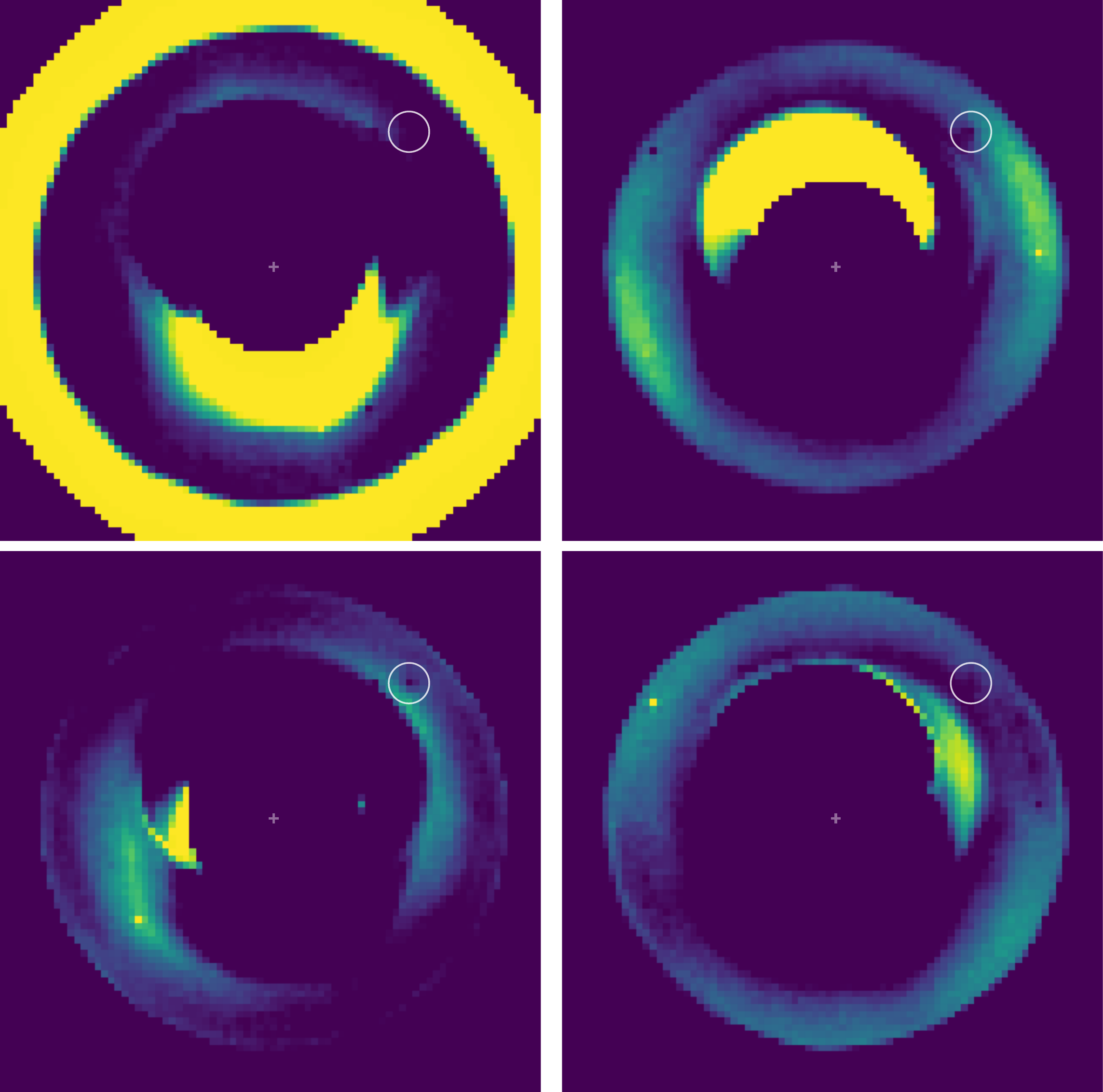}}
    \caption{Final result of post-processing the synthetic HARMONI data with a cross-correlation with the planet spectral template (left) and spectral unmixing through orthogonal subspace projection (middle and right). The number of spectral signatures is set to eight to decompose the data. The injected mock planet has a broadband contrast of $5\times10^{-6}$ with the central star  at a separation of 100 mas (white circle). Images are normalized to the peak value of the planet for each algorithm, the scale is linear, the field of view is identical, and the white cross marks the position of the star.}
    \label{fig:harmoni_test}
\end{figure*}

A test case was generated to give a first impression of our spectral unmixing algorithm. A simulated planet with a contrast of $5\times10^{-6}$ was injected at 100 mas and a position angle of $315^{\circ}$ into the simulated data cube. For each cube, the stellar halo was subtracted in the spectral direction in each data cube following \citet{Hoeijmakers:2018}, as suggested in section \ref{sec:maths}. The frames were then aligned north along the vertical axis and averaged over the temporal axis. The final 3D data cube was decomposed into eight sample  spectra and their corresponding spatial maps through spectral unmixing following the steps given in section \ref{sec:maths}. To compare the result with a spectrum-based detection algorithm, each spaxel in the final 3D data cube was cross-correlated with the injected model spectrum, whose pseudo-continuum was subtracted beforehand with a convolution with a Gaussian line-spread function corresponding to a spectral resolution of 100. The cross-correlation used a customized version of \texttt{CrosscorrRV}\footnote{\url{https://pyastronomy.readthedocs.io/en/latest/pyaslDoc/aslDoc/crosscorr.html}} to include normalization with the variance of the model and the data. A perfect match yields 1.

\begin{figure}
    \centering
    \includegraphics[width=0.5\textwidth]{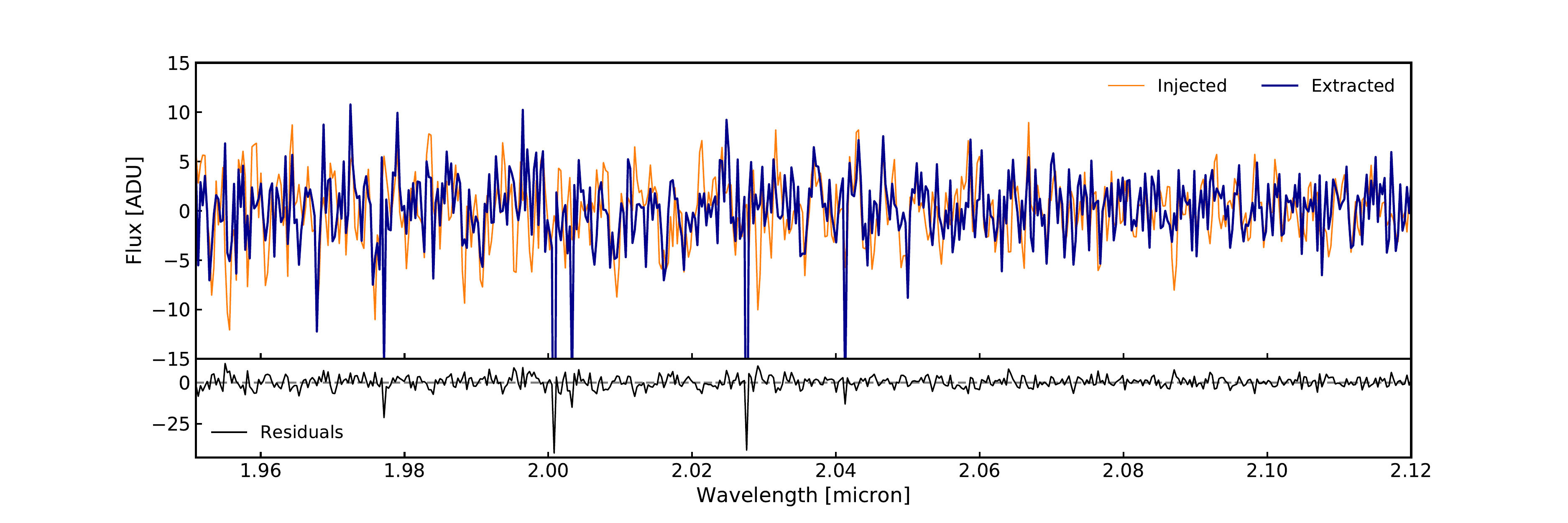}
        \includegraphics[width=0.5\textwidth]{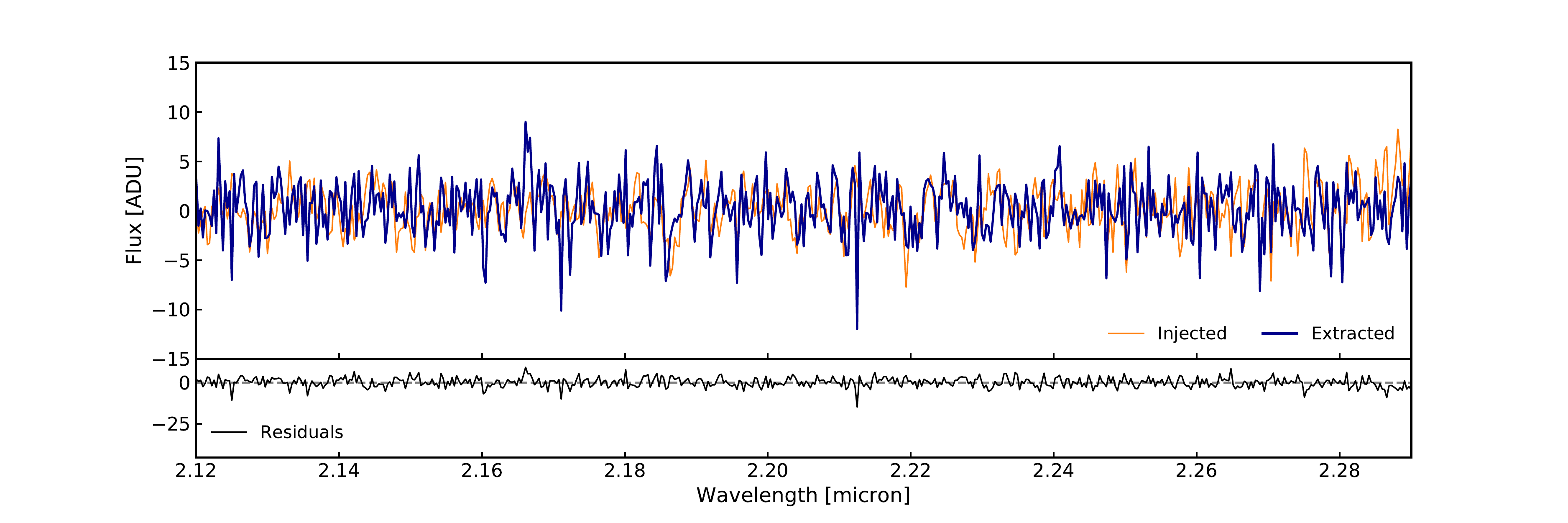}
            \includegraphics[width=0.5\textwidth]{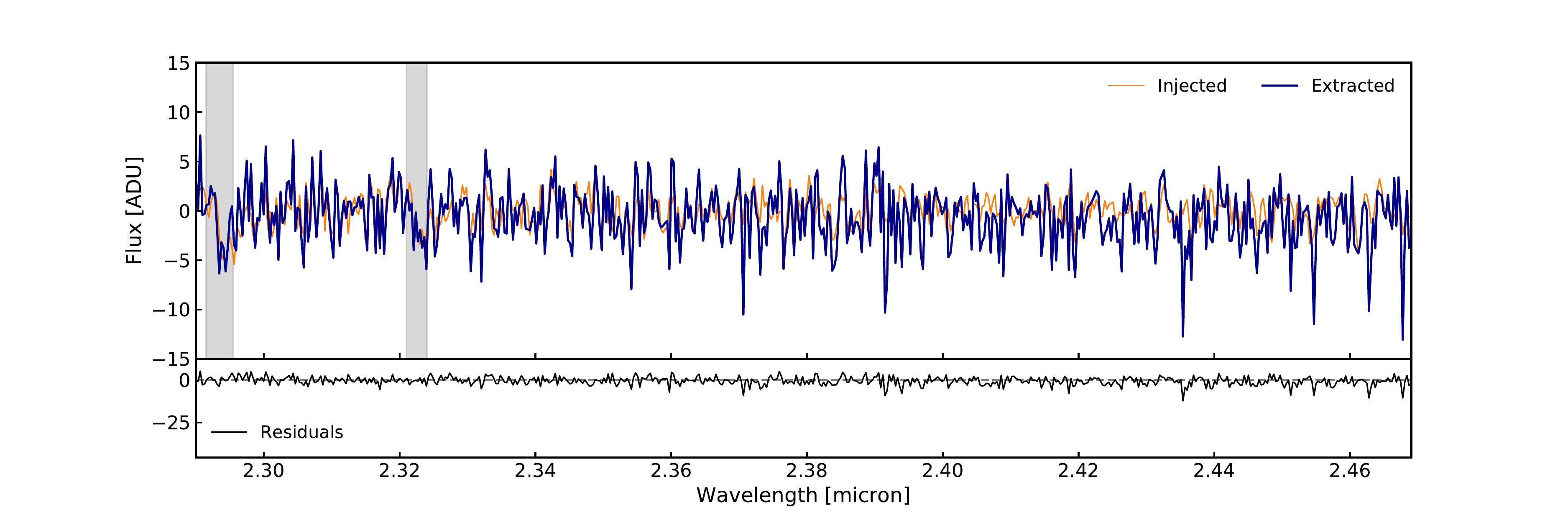}
         \includegraphics[width=0.5\textwidth]{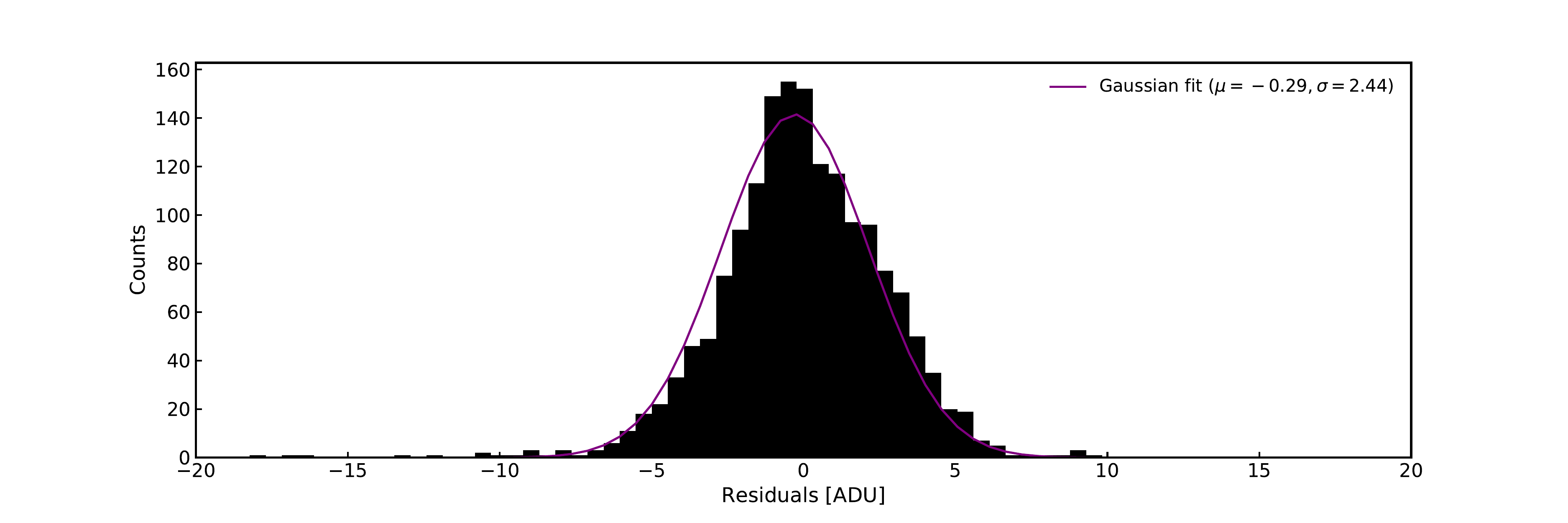}
    \caption{Inferred spectrum of the planet with a contrast $5\times10^{-6}$ injected into the synthetic HARMONI data (see Figure \ref{fig:harmoni_test} with spectral unmixing through orthogonal subspace projection (dark blue), compared with the high-frequency component of the injected spectrum (orange) along with the difference between the two (black). The first-overtone bandheads of \chem{CO} are highlighted as being well recovered (shade). The unmixing is not perfect, and the spectrum of the planet is contaminated with residuals from the subtraction of the diffracted starlight (bottom).}
    \label{fig:harmoni_test_spec}
\end{figure}

Figure \ref{fig:harmoni_test} shows the final images using the cross-correlation and spectral unmixing algorithms. Spectral unmixing can clearly separate the speckles from the planetary signal. The field of view is further decomposed into several regions that correspond to the variation of the stellar spectrum around the outer working angle, partial coverage of some pixels by the asymmetric focal plane mask after derotation and stack, or interpolation effects close to the mask (dominantly yellow regions). The pixels at the location of the planet have zero weight in the corresponding maps. We computed the S/N of the planet and the background pixels at the same separation, following the definition with small sample statistics correction of \citet{Mawet:2014}, to assess the detection. We used an aperture size of $2.6$ pixels, which is the average FWHM measured on the off-axis PSF. The S/N with the cross-correlation is 23.6, while the S/N with the spectral unmixing is 12.6. The distribution of the values in the cross-correlation and the weight maps are given in Appendix \ref{appendix}. If the decomposition were exact, all pixels in the background would have zero weights, which would cause the S/N to become infinite. Figure \ref{fig:harmoni_test} shows the extracted spectrum corresponding to the planet, compared with the injected model. The match is visually reasonable, in particular the bandheads of the first overtone of \chem{CO} at $2.295\,\micron$ and at $2.324\,\micron$, which are well recovered. The distribution of the difference between the two spectra as shown in the bottom panel of Figure \ref{fig:harmoni_test_spec} is Gaussian and almost centered on zero ($-0.28$). Residual noise from the subtraction of the diffracted halo contaminates the inferred spectrum of the planet by construction. These residuals are in common with that of the spaxels in the background at the same separation, explaining their small but nonzero weights. To further show that the extracted spectrum holds information about the true planet spectrum, the amplitude of the cross-correlation with the injected model is 0.15, which is close to the amplitude (0.17) at the planet location in the cross-correlation map, as expected.  Cross-correlating the $p-1$ nonplanetary spectra with the injected model yields $0.025\pm0.010$ on average. This confirms that the planetary information content is gathered in a single spectrum.

The cross-correlation leads to an S/N that is about twice as high as with spectral unmixing in this test case. However, 
the case is ideal for \textit{\textup{supervised}} detection techniques such as a cross-correlation because the searching template is the same as the injected one. Only contrast and noise matter. The lower S/N with the \textit{\textup{unsupervised}} spectral unmixing approach does not prevent the detection of the planet, and the extracted spectrum can be analyzed directly. In section \ref{sec:perf} we test whether this holds true with fainter planets. Nevertheless, spectral unmixing provides a prior-free alternative to classic cross-correlation.
%SNR(p=3)=5.8, SNR(p=25)=22.86

\section{Performance}
\label{sec:perf}

\subsection{Method}

In order to explore the properties of spectral unmixing in more detail and compare it with classic cross-correlation beyond a single test case, we computed the detection sensitivity in the central region, which is expressed in the star-to-planet contrast as a function of the separation. Simulated planets were injected from 68 to 132 mas with steps of 12 mas ($\sim1\,\lambda/D$) every $60\,\circ$ along a spiral. The injection was independently repeated six times by rotating the spiral by $60\circ$ increments. For each set of injections, the contrast was enhanced from $1\times10^{-7}$ to $1\times10^{-5}$, with steps of $5\times10^{-7}$ first and then refined with steps of $1\times10^{-7}$ around the limits. The detection limit is built upon the minimum contrast that yields 95\% of the injected simulated planets at a given separation, to be detected above a threshold of 5 in S/N \citep{Jensen:2018}. The process was repeated with spectral unmixing with $p=10$ and a cross-correlation with the same template.

\begin{figure}
    \centering
    \includegraphics[width=0.5\textwidth]{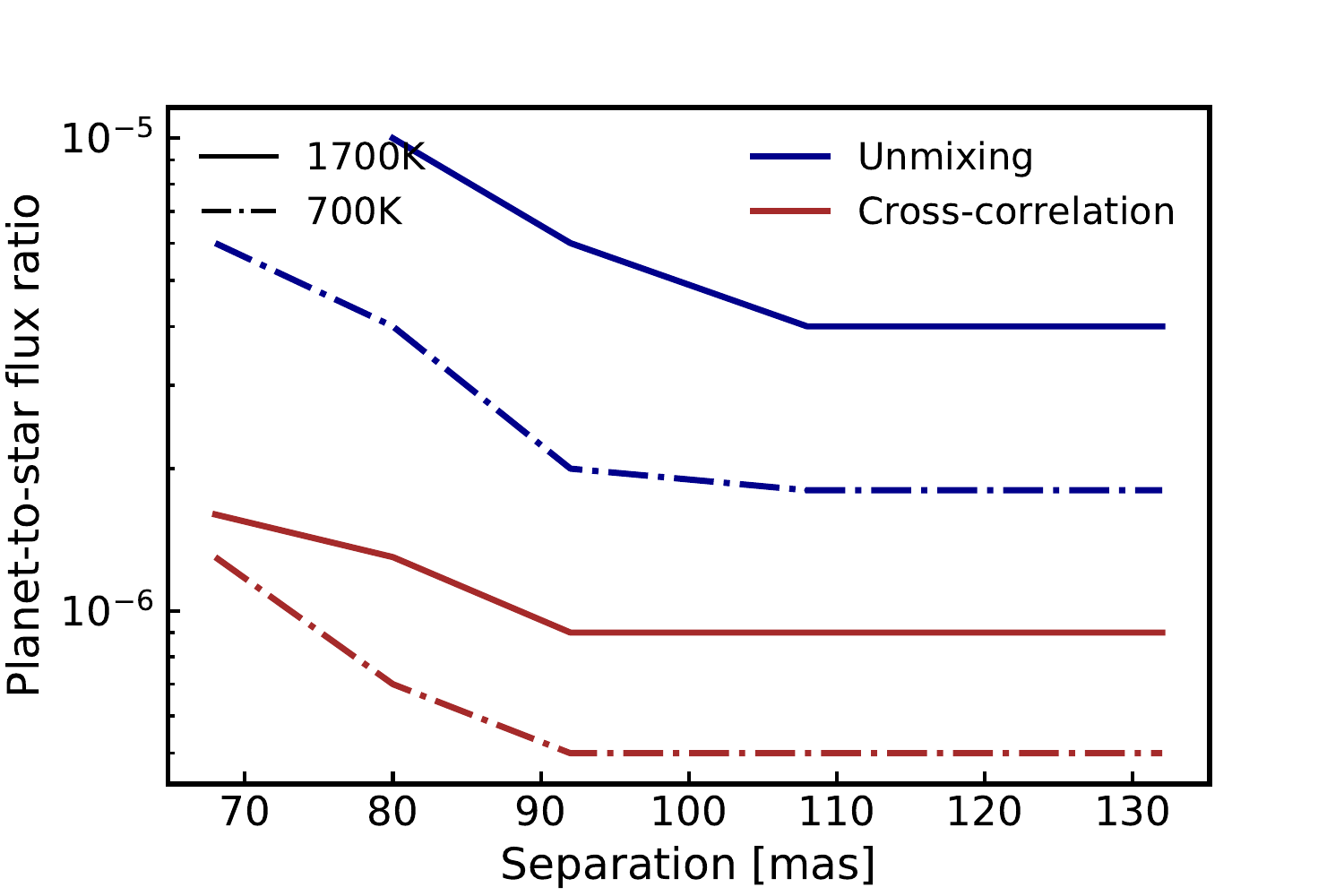}
    \caption{Completeness curves (95\%) with spectral unmixing (dark blue) and cross-correlation (brown) expressed as the minimum planet-to-star contrast for a detection threshold of $\tau=5\sigma$ as a function of the projected separation for L-type (solid) and T-type spectra (dot-dashed) in the synthetic HARMONI data.}
    \label{fig:HARMONI_contrast}
\end{figure}

\begin{figure}
    \centering
   \includegraphics[width=0.4\textwidth]{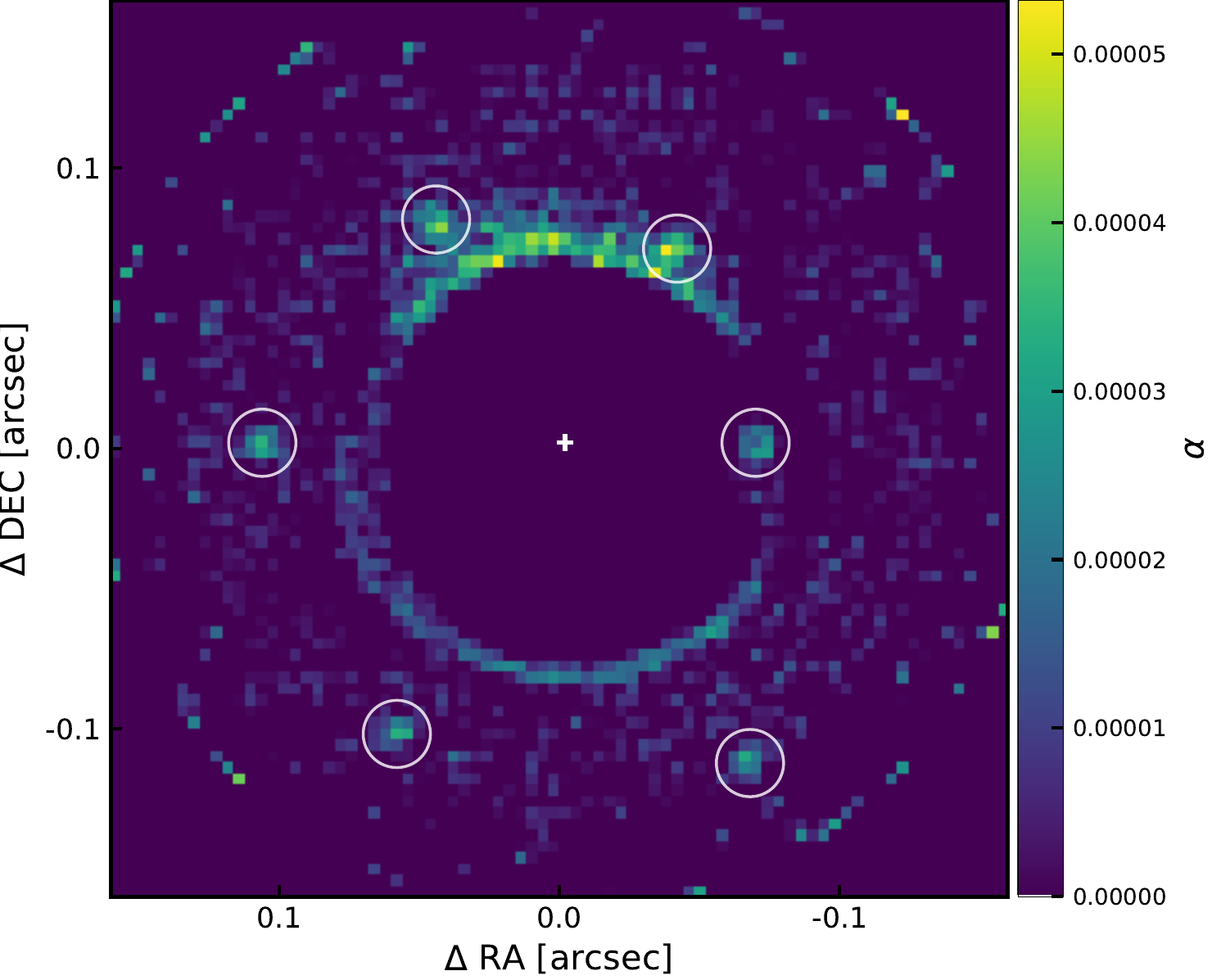}
    \includegraphics[width=0.4\textwidth]{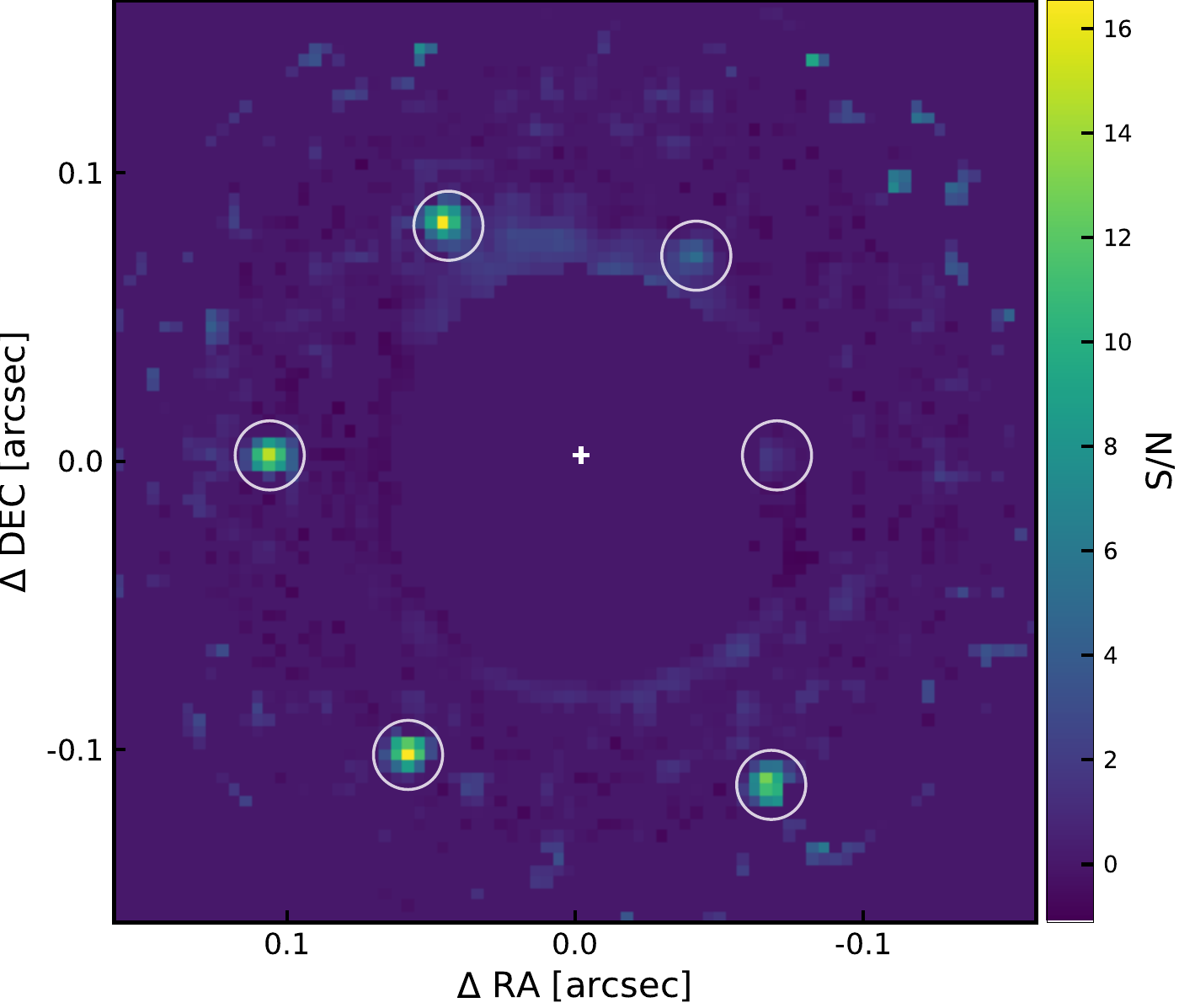}
\caption{Spatial weight inversion map of synthetic HARMONI data with a spiral of simulated planets (white circles) with a star-to-planet flux ratio of $9\times10^{-7}$ (top) and corresponding S/N map (bottom). The injected model spectrum is used to invert the data following equation \ref{eq:nnls}. Despite tiny spatial weights, most planets are detected with a high S/N.}
    \label{fig:HARMONI_faint}
\end{figure}

\subsection{Effect of  contrast and separation}
The 95\% completeness curves are shown in Figure \ref{fig:HARMONI_contrast}. Spectral unmixing is able to both flag planet-like spectra and lead to significant planet detections down to $4\times10^{-6}$ in this set of synthetic HARMONI data. 
The detection limit achieved with ideal supervised cross-correlation is more than four times deeper than with spectral unmixing at all separations. The question then arises whether the lower limit with spectral unmixing is reached because the S/N is below the adopted threshold or because the algorithm fails to glean the planet spectrum. The S/N  hits a hard floor around three down to a contrast of $3\times10^{-6}$. Fainter planets cannot be flagged in any of the spatial weight maps corresponding to the $p=10$ identified sample spectra; augmenting $p$ up to high values does not help. This critical step therefore limits the performance of spectral unmixing. To corroborate this, a data set with planets with a contrast $9\times10^{-7}$ was decomposed into $p=10$ samples. The planet model template was further added as an eleventh spectrum, and the data were inverted following equation \ref{eq:nnls}. The corresponding spatial weight map and its S/N map are displayed in Figure \ref{fig:HARMONI_faint}. All six planets can be spotted in the maps, with a high S/N for all but the two innermost ones. The weight corresponding to each planet is lower than $5\times10^{-3}\%$, however, meaning that these spaxels are highly mixed and that the planetary component is too weak to be extracted with equations \ref{eq:projector} and \ref{eq:distance}.

Spectral fidelity can also be investigated as a function of contrast. The planet spectrum as extracted with spectral umixing was cross-correlated with the injected template. The amplitude linearly decreases from 0.15 to to 0.07 with contrasts from $10^{-5}$ to $3\times10^{-6}$. This is more than three times than the amplitude of the cross-correlation of all other $10-1$ spectra with the planet template, which is stable around $0.027\pm0.005$. 
\subsection{Effect of the planet spectrum}
The previous results hold for the planet spectrum at 1700\,K, or L-type planets. However, changes in the intrinsic properties of the absorbing molecules (e.g., temperature, pressure, or abundance) alter the number and depth of the lines in the planet spectrum. As the S/N achieved with cross-correlation at first order scales with these two parameters \citep{Snellen:2015}, we might expect that they might also help pushing the limits with spectral unmixing.

We used a significantly different model spectrum with an effective temperature of 700\,K. This is illustrative of known T-type planets (e.g., 51\,Eri\,b). The atmosphere is now cool enough such that methane can form and produce many absorption lines, particularly in the K band, in addition to the lines due to water and carbone monoxyde. The injection and recovery process was repeated with this new template to build the 95\% completeness curves for spectral unmixing and the cross-correlation.

The resulting curves are shown in Figure \ref{fig:HARMONI_contrast}. Spectral unmixing can detect fainter T-type planets than L-type planets, the limiting contrast being improved by 55\,\% down to $1.8\times10^{-6}$. The sensitivity is also boosted with the cross-correlation, as predicted. T-type planets with contrast as faint as $5\times10^{-7}$ can be detected, which is an improvement of 45\,\% compared with L-type planets. A switch from L-type to T-type spectra yields a higher gain for spectral unmixing and thus reduces the gap between the two methods from $4.4$ to $3.6$. This difference might be even further reduced for real planets. Model spectra do not perfectly reproduce observed spectra, therefore template mismatch will degrade the performance of the cross-correlation, while spectral unmixing should remain unchanged. 

It has to be noted that the completeness curves presented here do not predict the ultimate sensitivity of HARMONI. They explore a very limited range of planet properties and post-processing algorithms. This is not the goal of the present work; more exhaustive results can be found in \citet{Houlle:2021}.

\section{Validation on real VLT/SINFONI data}
\label{sec:sinfoni}
The spectral unmixing algorithm was further vetted with real VLT/SINFONI data. Although SINFONI was not designed as a high-contrast instrument, it delivers medium-resolution hyperspectral data such that our method can be used.

\subsection{Data and preprocessing}

The data set of $\beta$ Pictoris and its planetary companion $\beta$ Pictoris b \citep{Lagrange:2010} obtained on September 10, 2014, was used (\citealt{Hoeijmakers:2018}, Bonnefoy et al. in prep). Observations were made in angular differential imaging mode at K band ($1.929-2.472\,\micron$, $R\sim5000$) with the high spatial resolution mode. This provided a $0.8\,"\times0.8\,"$ field of view with a scale of $12.5\times25\,\mathrm{mas}^2/\mathrm{pixel}$. The adaptive optics system was locked on the primary star, yielding a Strehl ratio of 17-29\% under median seeing conditions (0.7-0.9\,"). The star was kept outside the field of view during the sequence in order to prevent saturation. Twenty-four cubes with an integration time of 60s were obtained. Data cubes were initially built from the 2D raw detector images  with the SINFONI data-handling pipeline v.3.2.3 \citep{Abuter:2006}. The pipeline calibrated and corrected the instrument transmission and the distortion, identified and interpolated hot and nonlinear pixels, retrieved the position of the 32 dispersed slitlets in the raw frames, achieved the wavelength calibration, and finally built the 3D spatial-spectral cubes. We used the \texttt{TExTRIS} package (\citealt{Petrus:2020}, Bonnefoy et al., in prep.) prior to the ESO pipeline to correct the raw frames for various electronic noises. \texttt{TExTRIS} was also used to correct the data cubes for the improper estimates of the slitlet edges provided by the pipeline and for residual wavelength shifts. It computed the parallactic angles and retrieved the position of the star outside of the field of view. The diffracted starlight was then removed from each cube following \citet{Hoeijmakers:2018}. All cubes were aligned with north up and averaged at each wavelength slice to produce a single hyperspectral data cube.
Further details can be found in \citet{Petrus:2020} and Bonnefoy et al., (in prep).

\subsection{Results}

\begin{figure}
    \centering
    \includegraphics[width=0.4\textwidth]{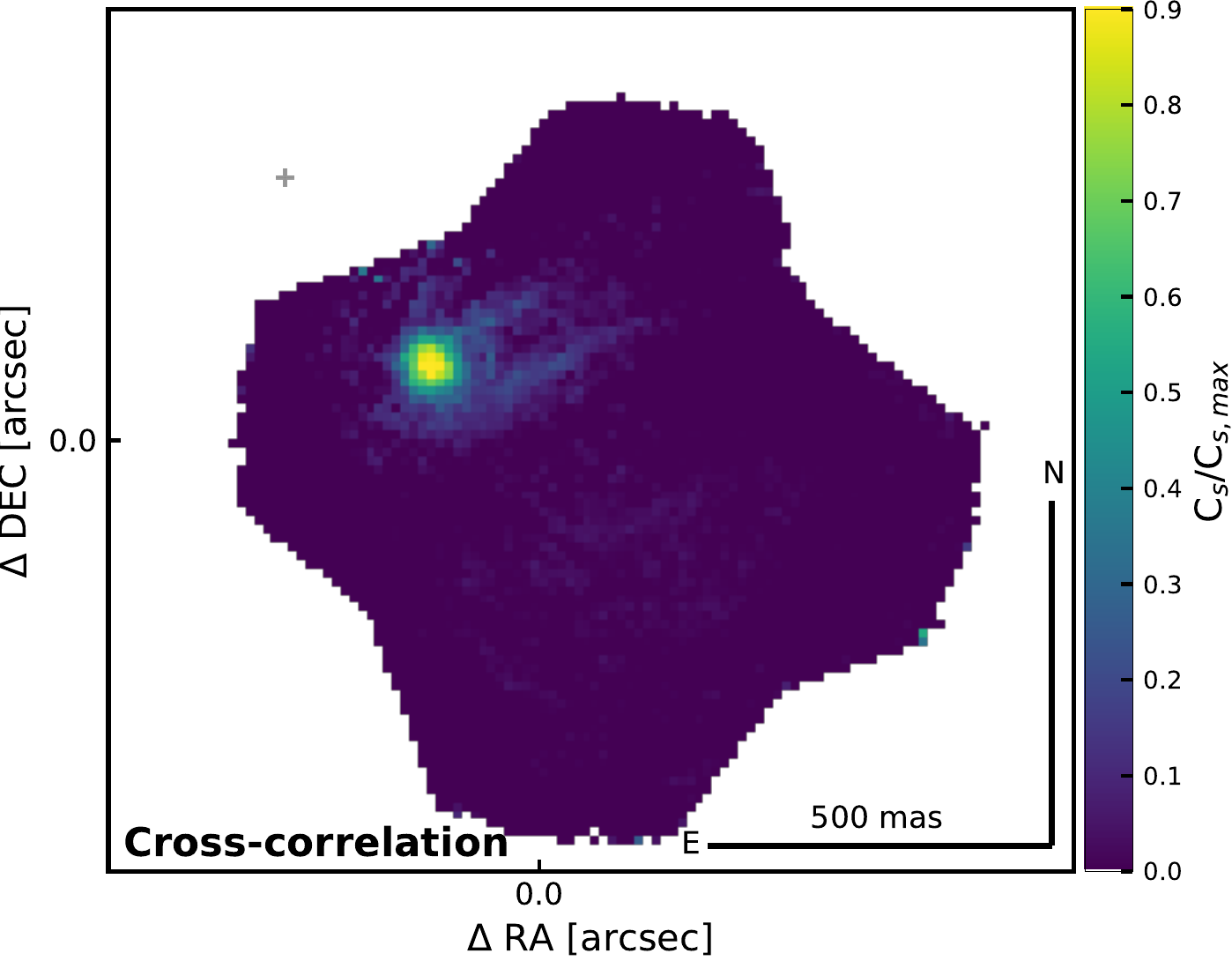}
    \includegraphics[width=0.4\textwidth]{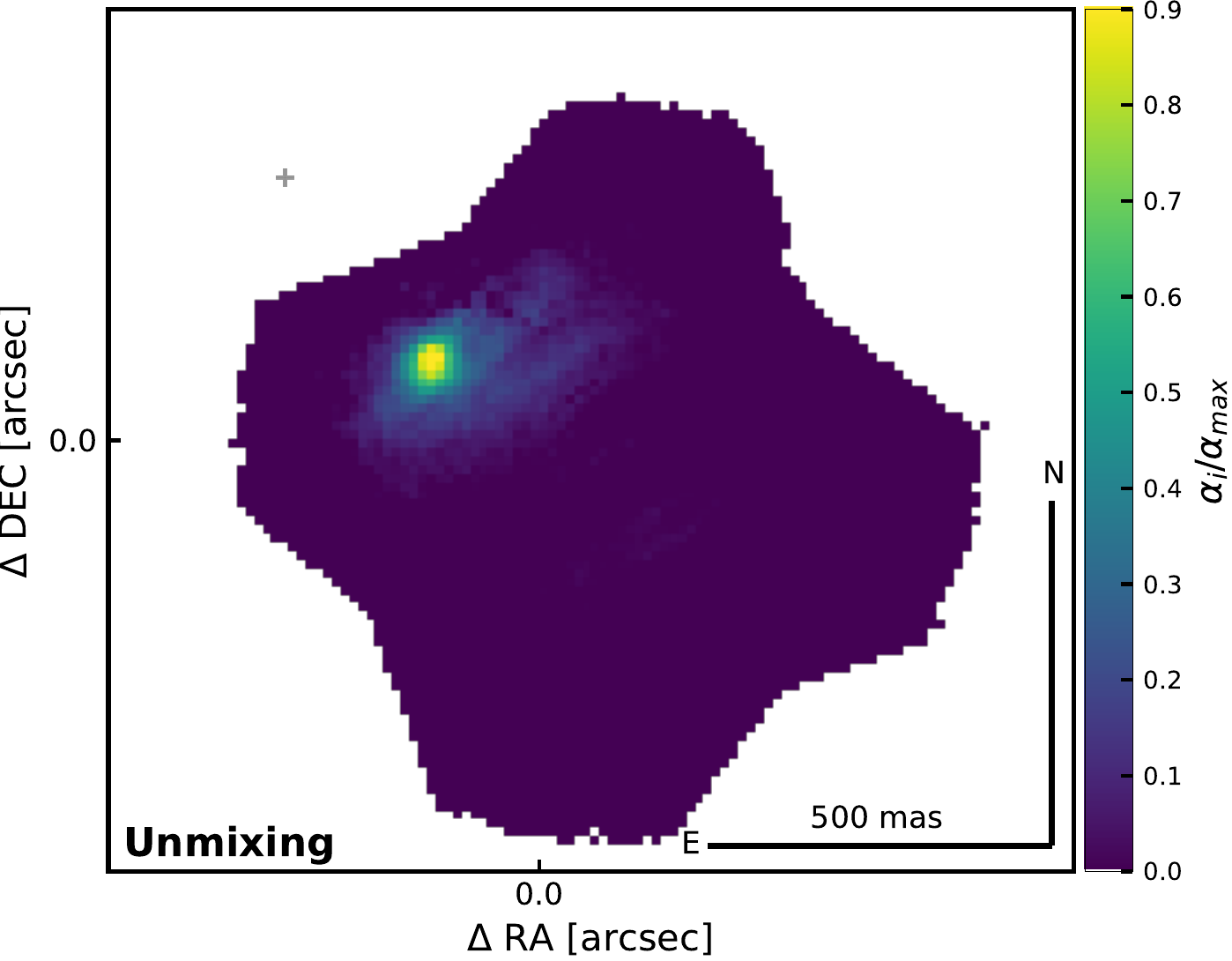}
    \caption{Final results of post-processing VLT/SINFONI K-band data of $\beta$ Pictoris b with the cross-correlation (top) and spectral unmixing through orthogonal subspace projection (bottom). A BT-Settl model spectrum (1700\,K, $\log g=4.0$) was used for the cross-correlation, and $p=10$ was set to decompose the data. The scale is linear and was chosen to reach 90\% of the peak value for each image, the cross-correlation strength $C_s$ , and the weight $\alpha_i$ , respectively. The cross marks the position of $\beta$ Pictoris A.}
    \label{fig:SINFONI_Bpic}
\end{figure}

$\beta$ Pictoris b \citep[known contrast of $2.1\times10^{-4}$, separation of $\sim$360mas at the time of the observations,][]{Bonnefoy:2011, 2019A&A...621L...8L} was searched for in the hyperspectral data cube with cross-correlation and spectral unmixing. The cross-correlation was performed over a range of radial velocities of $\pm100\,\mathrm{km}.\mathrm{s}^{-1}$ with respect to the star with steps of $\pm10\,\mathrm{km}.\mathrm{s}^{-1}$. The template spectrum was a continuum-removed BT-Settl model that matched the effective temperature ($\sim1700$\,K) and surface gravity ($\sim4.0$\,dex) of the planet \citep{Chilcote:2017,Nowak:2020}. For the spectral unmixing, the data were assumed to be decomposed into at least $p=10$ sample spectra, in the same way as for the HARMONI data. Figure \ref{fig:SINFONI_Bpic} shows the resulting images with the two algorithms. Spectral unmixing clearly unveils the planet, while the majority of the background pixels have zero weight with the planet spectrum. The CCF peaks at the planet position in the central velocity bin ($0\,\mathrm{km}.\mathrm{s}^{-1}$), as found by \citet{Hoeijmakers:2018}. Because the reconstructed field of view is asymmetrical, a pseudo-S/N was computed from the mean flux and standard deviation of all resolution elements in the whole field of view, with a diameter of 4 pixels. $\beta$ Pictoris b has a pseudo-S/N of 28.5 with spectral unmixing, and this value is 8.2 with the cross-correlation.

\begin{figure}
    \centering
    \includegraphics[width=0.5\textwidth]{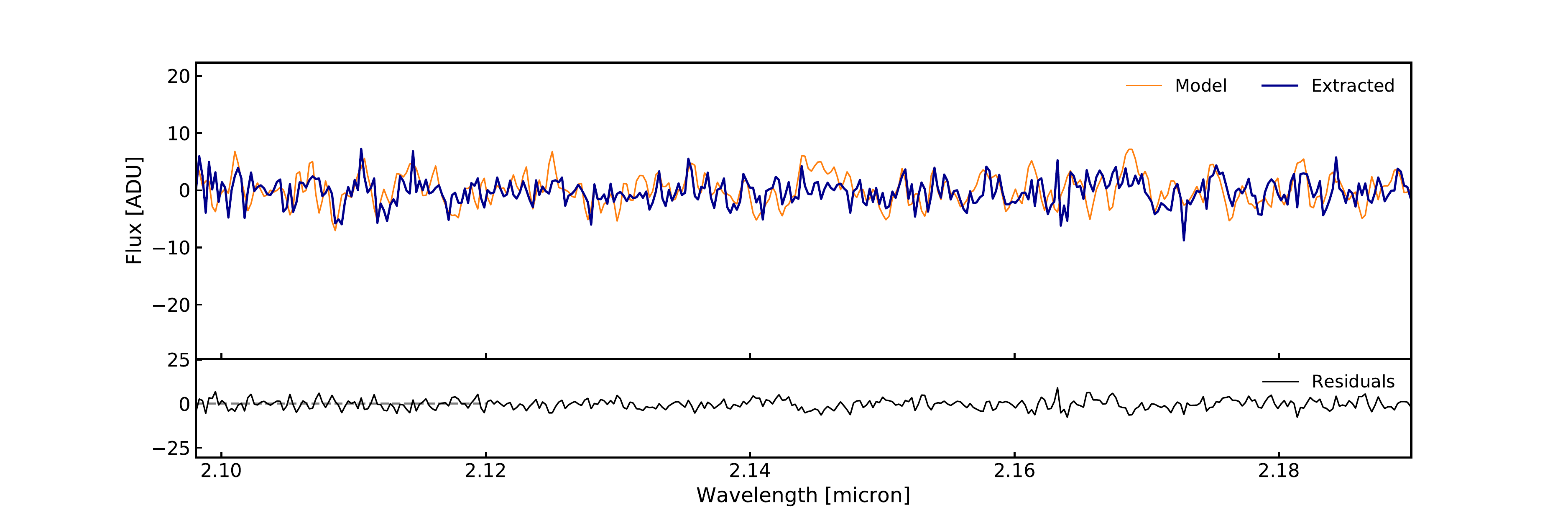}
        \includegraphics[width=0.5\textwidth]{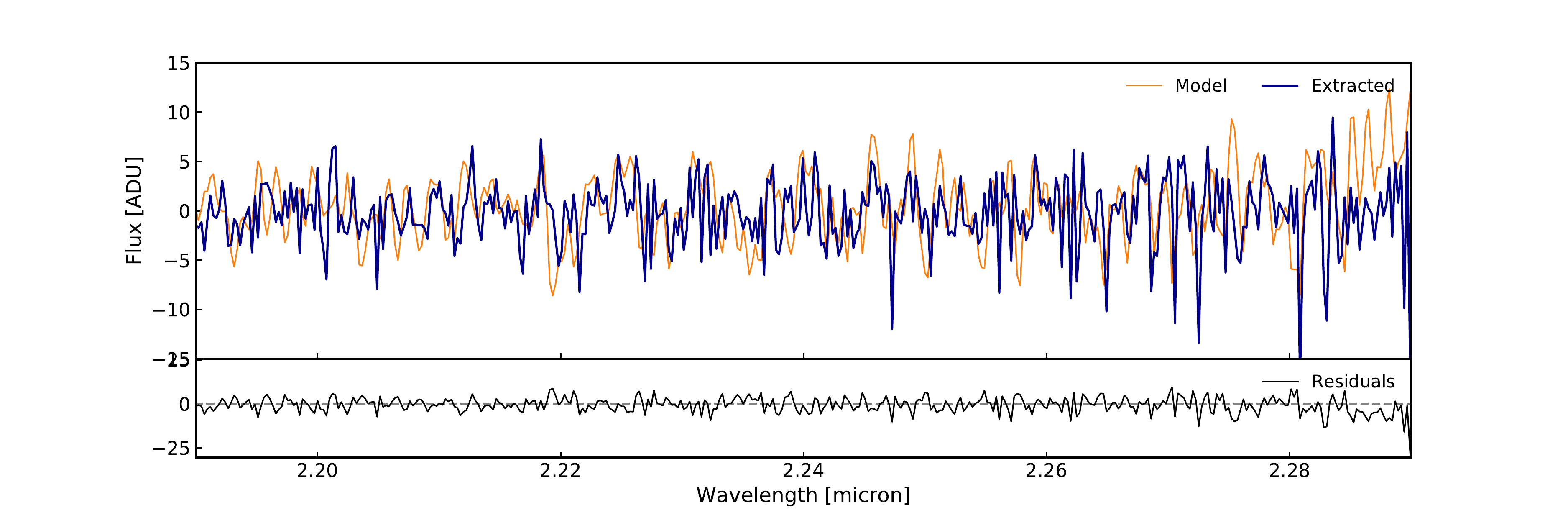}
    \includegraphics[width=0.5\textwidth]{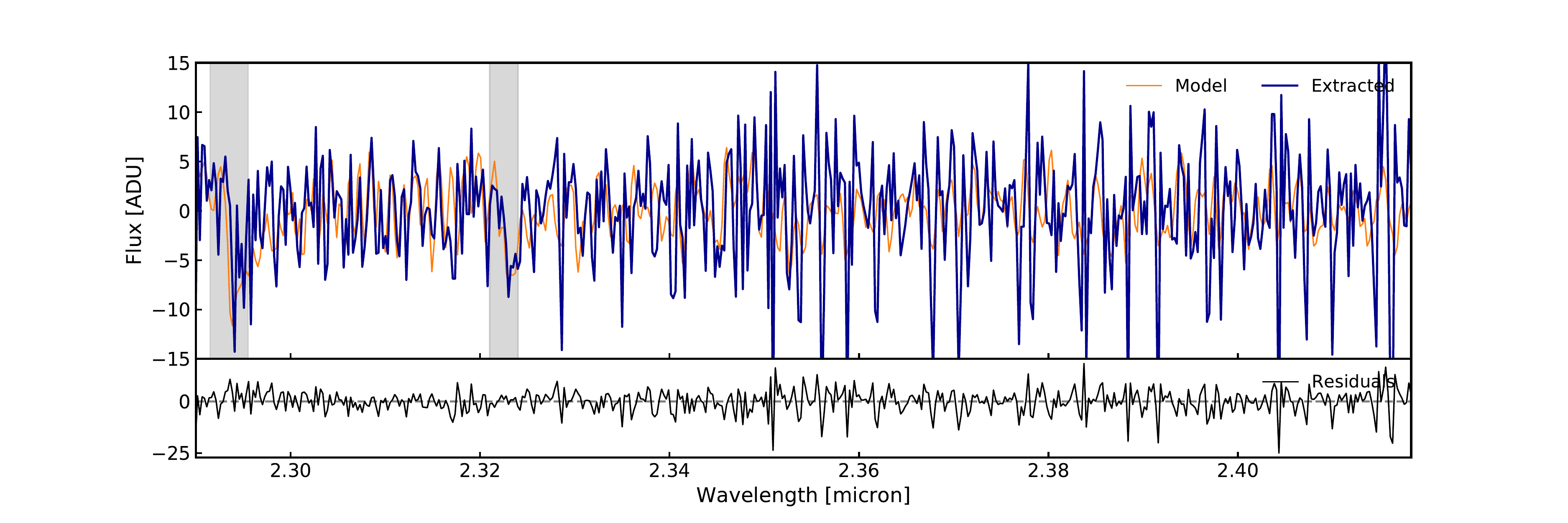}
    \includegraphics[width=0.5\textwidth]{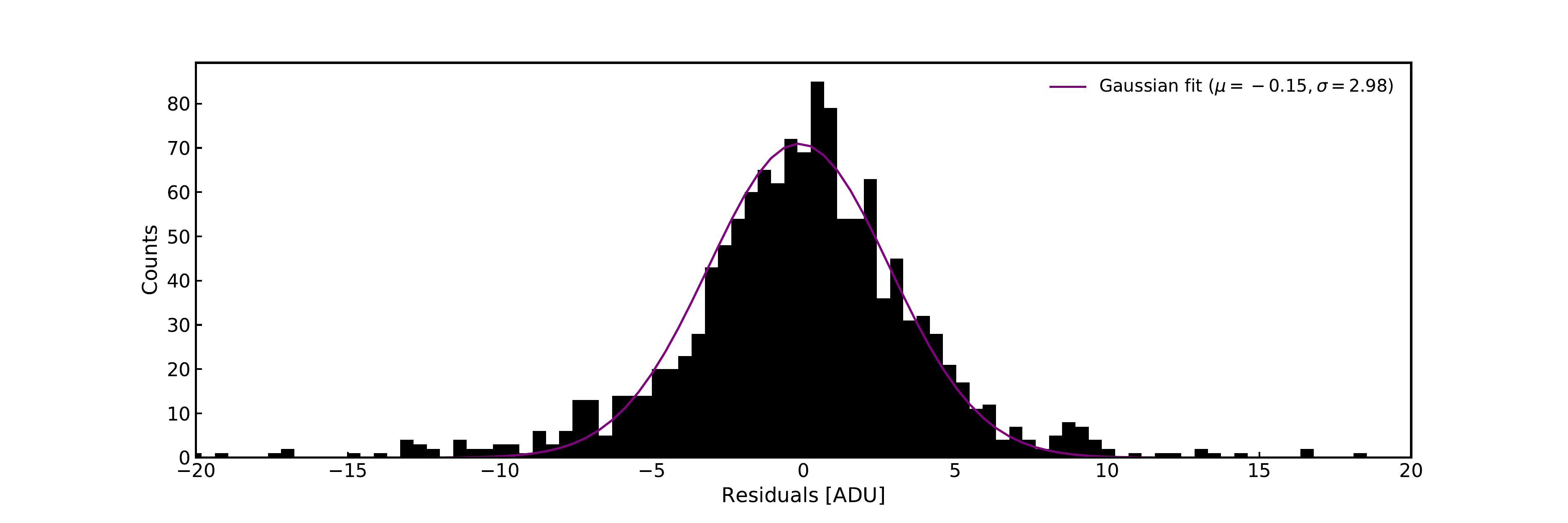}
    \caption{Top: Extracted continuum-removed SINFONI spectrum of $\beta$ Pictoris b with spectral unmixing (dark blue), compared with a BT-Settl model spectrum at 1700\,K and $\log g=4.0\,dex$ (orange), close to the true planet properties. The first-overtone bandheads of \chem{CO} are highlighted as being well recovered (shade). The extracted spectrum is not perfect, it is contaminated by unmixed residuals. Bottom: Histogram of the residuals (black) with a best-fit Gaussian function (purple).}
    \label{fig:bpic_spec}
\end{figure}

The spectrum corresponding to $\beta$ Pictoris b in the spatial weight map in Figure \ref{fig:SINFONI_Bpic} is displayed in Figure \ref{fig:bpic_spec}. A visual comparison can be established with the BT-Settl model spectrum. Although it is not a perfect model of the planet, several common lines can be spotted, including two bandheads of the first overtone of \chem{CO} at $2.295\,\micron$ and at $2.324\,\micron$. The histogram of the residuals does not exhibit a prominent bias ($\mu=-0.15$\,ADU), but shows a large spread ($\sigma=2.98$\,ADU). This is supported by the cross-correlation of the model spectrum with the extracted planet spectrum, which yields $0.33$ (the cross-correlation with the seven other extracted spectra gives $0.07\pm0.05$). However, the region beyond $2.34\,\micron$ is noticeably polluted by residuals that are not properly decomposed from the planet spectrum. Again, this is expected given the hypothesis of spectral unmixing, which is discussed in section \ref{sec:discussion}.

\section{Discussion and concluding remarks}
\label{sec:discussion}

\subsection{Limitations and benefits}
 Several limitations to spectral umixing have been identified throughout this work. First, the algorithm fails to detect a planet as faint as with the classic cross-correlation in synthetic data, which is favored by the injection and recovery process using the same model spectrum. However, as experimented on the real SINFONI data, for which it led to a higher S/N for $\beta$ Pictoris\,b, the sensitivity gap between the two methods may be reduced because the cross-correlation is affected by template mismatch. The extraction steps were shown to be responsible for this (see section \ref{sec:perf}). The orthogonal projection distance was used as a metric to quantify spectral dissimilarities between the tested spaxels. Alternative distances might be used. Cross-correlation and spectral angle mapper, a common metric in hyperspectral remote sensing, defined the angle between two vectors formed by two spectra \citep{Kruse:1993}, were tested but they led to worse performances (see Appendix \ref{ap:dist}). For faint planets, the spectrum of the planet has a very small weight on the corresponding spaxels compared to the residual diffracted starlight even after the continuum was removed. In hyperspectral remote sensing, they are commonly treated as highly mixed spaxels. The identification of sample spectra in this case is a very difficult task for which a geometry-based algorithm such as we adopted, is not robust enough \citep{Bioucas-Dias:2012}. As hypothesized in section \ref{sec:maths}, the algorithm flags the source sample spectra as the purest spaxels in the field of view, which are then used to decompose the remaining spaxels. With highly mixed spaxels at the planet location, the hypothesis is no longer valid to flag and unmix the planet. These phenomena also lead to a second limitation of this approach. As we showed in sections \ref{sec:perf} and \ref{sec:sinfoni}, the extracted spectrum that leads to the planet detection in the spatial weight maps, is not the true planet spectrum. The decomposition is not exact (the true value of p remains unknown), the diffracted starlight contaminates the spectrum. As a consequence, the true contrast of a planet cannot be obtained with spectral unmixing and needs another post-processing approach.

Beyond these limitations, the spectral unmixing algorithm presented in this work has two advantages. It is fully data-driven, or unsupervised, and computationally fast. Both characteristics make it interesting with respect to the other strategy for analyzing hyperspectral data, the classic cross-correlation. For a purely blind search of real data, the cross-correlation must rely on theoretical planet spectra and explore many parameters for the planet(s) that are to be searched for, including the effective temperature, the surface gravity, the metallicity and composition, and the radial velocity. This will become even more dramatic with the forthcoming ELT/HARMONI, METIS, and \textit{JWST}/MIRI, which should be sensitive to young Saturn- and Neptune-like planets as well as mature Jupiters. The spectral diversity of these populations of planets is poorly known and might deserve an exploration of even larger parameter space. Expressing a robust detection limit that encompasses all these possibilities might therefore be challenging and a template mismatch might lead to false negatives. While a single cross-correlation is computationally very fast, the process can take several minutes to hours for the whole field of view. For huge amounts of data such as those provided by the forthcoming HARMONI and METIS instruments and/or with large template libraries, the task quickly becomes computationally inefficient. This is not the case with spectral unmixing, which takes 15 seconds to process the 16 Gb of the synthetic HARMONI data on a standard laptop with a dual-processor core running at 2.4 GHz.

\subsection{Summary and future work}
To conclude, we have introduce a new strategy for analyzing high-contrast medium spectral resolution data aiming at directly detecting and characterizing exoplanets. We used the dissimilarities between the spectrum of a planet and that of the remaining field of view, which are mostly variations in the spectrum of the central star. The approach comes from remote-sensing research, in which the spectral diversity is a driver of algorithm development, namely all methods in the class of spectral unmixing. A classic algorithm based on orthogonal subspace projection and non-negative least-squares inversion was proposed. It was demonstrated to be viable to simultaneously detect a planet and extract spectral information content in synthetic and real hyperspectral data. This approach is fully data-driven as opposed to a cross-correlation that relies on theoretical templates, which is commonly adopted to analyze such data. The unsupervised nature of spectral unmixing comes at the price of a reduced sensitivity compared to the supervised cross-correlation in the ideal case of synthetic data with the adopted stellar and planet models, but this may not be the case for different settings and real data. Nevertheless, the algorithm benefits from low computational cost, making it suitable for real-time processing in a purely blind search scenario.

The spectral unmixing algorithm presented in this work is solely based on spectral dissimilarities between spaxels in the field of view. This basic assessment is enough to demonstrate the potential of this class of strategies in decomposing hyperspectral data aiming at simultaneously detecting and characterizing exoplanets. Additional constraints and more expensive formulations of the decomposition based on the knowledge of the data are possible. Spectral umixing is a very active field of research, therefore we point out only a few directions of high interest. Spatial regularizer, taking into account the size of the PSF, can be added because nearby pixels share a similar spectral signature, and the decomposition of neighboring pixels might be forced to be comparable for the same sample spectra \citep[e.g.,][]{Zhang:2018}. Furthermore, the planet signal is sparse in spatial and spectral dimensions, while the remaining field of view is low dimensional in the spectral direction. Low-rank and sparse unmixing can thus be foreseen \citep[e.g.,][]{Tsinos:2017,Xu:2018}, similar to the reasoning behind the decomposition of angular differential imaging data from \citet{Gomez:2016}. The sparsity constrain can be pushed even further. The planet can be considered as a spectral anomaly in the field of view, or target, and anything else may be considered as background \citep[e.g.,][]{Li:2015,Soofbaf:2018,Yang:2019}. These improvements could lead to better sensitivity and unbiased spectral extraction. They will be the focus of future work.

\begin{acknowledgements}
      JR is supported by the French National Research Agency in the framework of the Investissements d’Avenir program (ANR- 15-IDEX-02), through the funding of the "Origin of Life" project of the Univ. Grenoble-Alpes. AC acknowledges support from the European Research Council (ERC) under the European Union’s Horizon 2020 research and innovation program, for the Project "EXACT".
\end{acknowledgements}

   \bibliographystyle{aa} % style aa.bst
   \bibliography{biblio.bib} % your references Yourfile.bib

\begin{thebibliography}{114}
\expandafter\ifx\csname natexlab\endcsname\relax\def\natexlab#1{#1}\fi

\bibitem[{{Abuter} {et~al.}(2006){Abuter}, {Schreiber}, {Eisenhauer}, {Ott},
  {Horrobin}, \& {Gillesen}}]{Abuter:2006}
{Abuter}, R., {Schreiber}, J., {Eisenhauer}, F., {et~al.} 2006, \nar, 50, 398

\bibitem[{{Acito} {et~al.}(2009){Acito}, {Diani}, \& {Corsini}}]{Acito:2009}
{Acito}, N., {Diani}, M., \& {Corsini}, G. 2009, IEEE Transactions on
  Geoscience and Remote Sensing, 47, 3844

\bibitem[{{Allard}(2014)}]{Allard:2014}
{Allard}, F. 2014, in Exploring the Formation and Evolution of Planetary
  Systems, ed. M.~{Booth}, B.~C. {Matthews}, \& J.~R. {Graham}, Vol. 299,
  271--272

\bibitem[{{Allard} {et~al.}(2012){Allard}, {Homeier}, \&
  {Freytag}}]{Allard:2012}
{Allard}, F., {Homeier}, D., \& {Freytag}, B. 2012, Philosophical Transactions
  of the Royal Society of London Series A, 370, 2765

\bibitem[{{Amara} \& {Quanz}(2012)}]{Amara:2012}
{Amara}, A. \& {Quanz}, S.~P. 2012, \mnras, 427, 948

\bibitem[{{Bagnasco} {et~al.}(2007){Bagnasco}, {Kolm}, {Ferruit}, {Honnen},
  {Koehler}, {Lemke}, {Maschmann}, {Melf}, {Noyer}, {Rumler}, {Salvignol},
  {Strada}, \& {Te Plate}}]{Bagnasco:2007}
{Bagnasco}, G., {Kolm}, M., {Ferruit}, P., {et~al.} 2007, in Society of
  Photo-Optical Instrumentation Engineers (SPIE) Conference Series, Vol. 6692,
  Cryogenic Optical Systems and Instruments XII, ed. J.~B. {Heaney} \& L.~G.
  {Burriesci}, 66920M

\bibitem[{{Barman} {et~al.}(2015){Barman}, {Konopacky}, {Macintosh}, \&
  {Marois}}]{Barman:2015}
{Barman}, T.~S., {Konopacky}, Q.~M., {Macintosh}, B., \& {Marois}, C. 2015,
  \apj, 804, 61

\bibitem[{{Barman} {et~al.}(2011){Barman}, {Macintosh}, {Konopacky}, \&
  {Marois}}]{Barman:2011}
{Barman}, T.~S., {Macintosh}, B., {Konopacky}, Q.~M., \& {Marois}, C. 2011,
  \apj, 733, 65

\bibitem[{{Bern{\'e}} {et~al.}(2007){Bern{\'e}}, {Joblin}, {Deville}, {Smith},
  {Rapacioli}, {Bernard}, {Thomas}, {Reach}, \& {Abergel}}]{Berne:2007}
{Bern{\'e}}, O., {Joblin}, C., {Deville}, Y., {et~al.} 2007, \aap, 469, 575

\bibitem[{{Beuzit} {et~al.}(2019){Beuzit}, {Vigan}, {Mouillet}, {Dohlen},
  {Gratton}, {Boccaletti}, {Sauvage}, {Schmid}, {Langlois}, {Petit},
  {Baruffolo}, {Feldt}, {Milli}, {Wahhaj}, {Abe}, {Anselmi}, {Antichi},
  {Barette}, {Baudrand}, {Baudoz}, {Bazzon}, {Bernardi}, {Blanchard}, {Brast},
  {Bruno}, {Buey}, {Carbillet}, {Carle}, {Cascone}, {Chapron}, {Charton},
  {Chauvin}, {Claudi}, {Costille}, {De Caprio}, {de Boer}, {Delboulb{\'e}},
  {Desidera}, {Dominik}, {Downing}, {Dupuis}, {Fabron}, {Fantinel}, {Farisato},
  {Feautrier}, {Fedrigo}, {Fusco}, {Gigan}, {Ginski}, {Girard}, {Giro},
  {Gisler}, {Gluck}, {Gry}, {Henning}, {Hubin}, {Hugot}, {Incorvaia}, {Jaquet},
  {Kasper}, {Lagadec}, {Lagrange}, {Le Coroller}, {Le Mignant}, {Le Ruyet},
  {Lessio}, {Lizon}, {Llored}, {Lundin}, {Madec}, {Magnard}, {Marteaud},
  {Martinez}, {Maurel}, {M{\'e}nard}, {Mesa}, {M{\"o}ller-Nilsson}, {Moulin},
  {Moutou}, {Orign{\'e}}, {Parisot}, {Pavlov}, {Perret}, {Pragt}, {Puget},
  {Rabou}, {Ramos}, {Reess}, {Rigal}, {Rochat}, {Roelfsema}, {Rousset}, {Roux},
  {Saisse}, {Salasnich}, {Santambrogio}, {Scuderi}, {Segransan}, {Sevin},
  {Siebenmorgen}, {Soenke}, {Stadler}, {Suarez}, {Tiph{\`e}ne}, {Turatto},
  {Udry}, {Vakili}, {Waters}, {Weber}, {Wildi}, {Zins}, \&
  {Zurlo}}]{Beuzit:2019}
{Beuzit}, J.~L., {Vigan}, A., {Mouillet}, D., {et~al.} 2019, \aap, 631, A155

\bibitem[{Bioucas-Dias \& Nascimento(2008)}]{Bioucas-Dias:2008}
Bioucas-Dias, J. \& Nascimento, J. 2008, Geoscience and Remote Sensing, IEEE
  Transactions on, 46, 2435

\bibitem[{{Bioucas-Dias} {et~al.}(2012{\natexlab{a}}){Bioucas-Dias}, {Plaza},
  {Dobigeon}, {Parente}, {Du}, {Gader}, \& {Chanussot}}]{Bioucas-Dias:2012}
{Bioucas-Dias}, J.~M., {Plaza}, A., {Dobigeon}, N., {et~al.}
  2012{\natexlab{a}}, IEEE Journal of Selected Topics in Applied Earth
  Observations and Remote Sensing, 5, 354

\bibitem[{{Bioucas-Dias} {et~al.}(2012{\natexlab{b}}){Bioucas-Dias}, {Plaza},
  {Dobigeon}, {Parente}, {Du}, {Gader}, \& {Chanussot}}]{Bioucas-Dias:2015}
{Bioucas-Dias}, J.~M., {Plaza}, A., {Dobigeon}, N., {et~al.}
  2012{\natexlab{b}}, IEEE Journal of Selected Topics in Applied Earth
  Observations and Remote Sensing, 5, 354

\bibitem[{{Bonnefoy} {et~al.}(2011){Bonnefoy}, {Lagrange}, {Boccaletti},
  {Chauvin}, {Apai}, {Allard}, {Ehrenreich}, {Girard}, {Mouillet}, {Rouan},
  {Gratadour}, \& {Kasper}}]{Bonnefoy:2011}
{Bonnefoy}, M., {Lagrange}, A.~M., {Boccaletti}, A., {et~al.} 2011, \aap, 528,
  L15

\bibitem[{{Bonnefoy} {et~al.}(2018){Bonnefoy}, {Perraut}, {Lagrange},
  {Delorme}, {Vigan}, {Line}, {Rodet}, {Ginski}, {Mourard}, {Marleau},
  {Samland}, {Tremblin}, {Ligi}, {Cantalloube}, {Molli{\`e}re}, {Charnay},
  {Kuzuhara}, {Janson}, {Morley}, {Homeier}, {D'Orazi}, {Klahr}, {Mordasini},
  {Lavie}, {Baudino}, {Beust}, {Peretti}, {Musso Bartucci}, {Mesa},
  {B{\'e}zard}, {Boccaletti}, {Galicher}, {Hagelberg}, {Desidera}, {Biller},
  {Maire}, {Allard}, {Borgniet}, {Lannier}, {Meunier}, {Desort}, {Alecian},
  {Chauvin}, {Langlois}, {Henning}, {Mugnier}, {Mouillet}, {Gratton}, {Brandt},
  {Mc Elwain}, {Beuzit}, {Tamura}, {Hori}, {Brandner}, {Buenzli}, {Cheetham},
  {Cudel}, {Feldt}, {Kasper}, {Keppler}, {Kopytova}, {Meyer}, {Perrot},
  {Rouan}, {Salter}, {Schmidt}, {Sissa}, {Zurlo}, {Wildi}, {Blanchard}, {De
  Caprio}, {Delboulb{\'e}}, {Maurel}, {Moulin}, {Pavlov}, {Rabou}, {Ramos},
  {Roelfsema}, {Rousset}, {Stadler}, {Rigal}, \& {Weber}}]{Bonnefoy:2018}
{Bonnefoy}, M., {Perraut}, K., {Lagrange}, A.~M., {et~al.} 2018, \aap, 618, A63

\bibitem[{{Bonnefoy} {et~al.}(2016){Bonnefoy}, {Zurlo}, {Baudino}, {Lucas},
  {Mesa}, {Maire}, {Vigan}, {Galicher}, {Homeier}, {Marocco}, {Gratton},
  {Chauvin}, {Allard}, {Desidera}, {Kasper}, {Moutou}, {Lagrange}, {Antichi},
  {Baruffolo}, {Baudrand}, {Beuzit}, {Boccaletti}, {Cantalloube}, {Carbillet},
  {Charton}, {Claudi}, {Costille}, {Dohlen}, {Dominik}, {Fantinel},
  {Feautrier}, {Feldt}, {Fusco}, {Gigan}, {Girard}, {Gluck}, {Gry}, {Henning},
  {Janson}, {Langlois}, {Madec}, {Magnard}, {Maurel}, {Mawet}, {Meyer},
  {Milli}, {Moeller-Nilsson}, {Mouillet}, {Pavlov}, {Perret}, {Pujet}, {Quanz},
  {Rochat}, {Rousset}, {Roux}, {Salasnich}, {Salter}, {Sauvage}, {Schmid},
  {Sevin}, {Soenke}, {Stadler}, {Turatto}, {Udry}, {Vakili}, {Wahhaj}, \&
  {Wildi}}]{Bonnefoy:2016}
{Bonnefoy}, M., {Zurlo}, A., {Baudino}, J.~L., {et~al.} 2016, \aap, 587, A58

\bibitem[{{Boulais} {et~al.}(2020){Boulais}, {Bern{\'e}}, {Faury}, \&
  {Deville}}]{Boulais:2020}
{Boulais}, A., {Bern{\'e}}, O., {Faury}, G., \& {Deville}, Y. 2020, arXiv
  e-prints, arXiv:2011.09742

\bibitem[{{Brandl} {et~al.}(2018){Brandl}, {Absil}, {Ag{\'o}cs}, {Baccichet},
  {Bertram}, {Bettonvil}, {van Boekel}, {Burtscher}, {van Dishoeck}, {Feldt},
  {Garcia}, {Glasse}, {Glauser}, {G{\"u}del}, {Haupt}, {Kenworthy}, {Labadie},
  {Laun}, {Lesman}, {Pantin}, {Quanz}, {Snellen}, {Siebenmorgen}, \& {van
  Winckel}}]{Brandl:2018}
{Brandl}, B.~R., {Absil}, O., {Ag{\'o}cs}, T., {et~al.} 2018, in Society of
  Photo-Optical Instrumentation Engineers (SPIE) Conference Series, Vol. 10702,
  Ground-based and Airborne Instrumentation for Astronomy VII, ed. C.~J.
  {Evans}, L.~{Simard}, \& H.~{Takami}, 107021U

\bibitem[{Brandt {et~al.}(2017)Brandt, Rizzo, Groff, Chilcote, Greco, Kasdin,
  Limbach, Galvin, Loomis, Knapp, McElwain, Jovanovic, Currie, Mede, Tamura,
  Takato, \& Hayashi}]{Brandt:2017}
Brandt, T., Rizzo, M., Groff, T., {et~al.} 2017

\bibitem[{Bro \& De~Jong(1997)}]{Bro:1997}
Bro, R. \& De~Jong, S. 1997, Journal of Chemometrics, 11, 393

\bibitem[{{Bryan} {et~al.}(2018){Bryan}, {Benneke}, {Knutson}, {Batygin}, \&
  {Bowler}}]{Bryan:2018}
{Bryan}, M.~L., {Benneke}, B., {Knutson}, H.~A., {Batygin}, K., \& {Bowler},
  B.~P. 2018, Nature Astronomy, 2, 138

\bibitem[{{Cantalloube} {et~al.}(2015){Cantalloube}, {Mouillet}, {Mugnier},
  {Milli}, {Absil}, {Gomez Gonzalez}, {Chauvin}, {Beuzit}, \&
  {Cornia}}]{Cantalloube:2015}
{Cantalloube}, F., {Mouillet}, D., {Mugnier}, L.~M., {et~al.} 2015, \aap, 582,
  A89

\bibitem[{{Carlotti} {et~al.}(2018){Carlotti}, {H{\'e}nault}, {Dohlen},
  {Sauvage}, {Rabou}, {Magnard}, {Vigan}, {Mouillet}, {Chauvin}, {Vola},
  {Bonnefoy}, {Fusco}, {El Hadi}, {Thatte}, {Clarke}, {Tecza}, {Bryson},
  {Schnetler}, \& {V{\'e}rinaud}}]{Carlotti:2018}
{Carlotti}, A., {H{\'e}nault}, F., {Dohlen}, K., {et~al.} 2018, in Society of
  Photo-Optical Instrumentation Engineers (SPIE) Conference Series, Vol. 10702,
  Ground-based and Airborne Instrumentation for Astronomy VII, ed. C.~J.
  {Evans}, L.~{Simard}, \& H.~{Takami}, 107029N

\bibitem[{Chang \& Du(2004)}]{Chang:2006}
Chang, C.-I. \& Du, Q. 2004, Geoscience and Remote Sensing, IEEE Transactions
  on, 42, 608

\bibitem[{{Chauvin} {et~al.}(2017){Chauvin}, {Desidera}, {Lagrange}, {Vigan},
  {Gratton}, {Langlois}, {Bonnefoy}, {Beuzit}, {Feldt}, {Mouillet}, {Meyer},
  {Cheetham}, {Biller}, {Boccaletti}, {D'Orazi}, {Galicher}, {Hagelberg},
  {Maire}, {Mesa}, {Olofsson}, {Samland}, {Schmidt}, {Sissa}, {Bonavita},
  {Charnay}, {Cudel}, {Daemgen}, {Delorme}, {Janin-Potiron}, {Janson},
  {Keppler}, {Le Coroller}, {Ligi}, {Marleau}, {Messina}, {Molli{\`e}re},
  {Mordasini}, {M{\"u}ller}, {Peretti}, {Perrot}, {Rodet}, {Rouan}, {Zurlo},
  {Dominik}, {Henning}, {Menard}, {Schmid}, {Turatto}, {Udry}, {Vakili}, {Abe},
  {Antichi}, {Baruffolo}, {Baudoz}, {Baudrand}, {Blanchard}, {Bazzon}, {Buey},
  {Carbillet}, {Carle}, {Charton}, {Cascone}, {Claudi}, {Costille}, {Deboulbe},
  {De Caprio}, {Dohlen}, {Fantinel}, {Feautrier}, {Fusco}, {Gigan}, {Giro},
  {Gisler}, {Gluck}, {Hubin}, {Hugot}, {Jaquet}, {Kasper}, {Madec}, {Magnard},
  {Martinez}, {Maurel}, {Le Mignant}, {M{\"o}ller-Nilsson}, {Llored}, {Moulin},
  {Orign{\'e}}, {Pavlov}, {Perret}, {Petit}, {Pragt}, {Puget}, {Rabou},
  {Ramos}, {Rigal}, {Rochat}, {Roelfsema}, {Rousset}, {Roux}, {Salasnich},
  {Sauvage}, {Sevin}, {Soenke}, {Stadler}, {Suarez}, {Weber}, {Wildi},
  {Antoniucci}, {Augereau}, {Baudino}, {Brandner}, {Engler}, {Girard}, {Gry},
  {Kral}, {Kopytova}, {Lagadec}, {Milli}, {Moutou}, {Schlieder},
  {Szul{\'a}gyi}, {Thalmann}, \& {Wahhaj}}]{Chauvin:2017}
{Chauvin}, G., {Desidera}, S., {Lagrange}, A.~M., {et~al.} 2017, \aap, 605, L9

\bibitem[{{Chauvin} {et~al.}(2018){Chauvin}, {Gratton}, {Bonnefoy}, {Lagrange},
  {de Boer}, {Vigan}, {Beust}, {Lazzoni}, {Boccaletti}, {Galicher}, {Desidera},
  {Delorme}, {Keppler}, {Lannier}, {Maire}, {Mesa}, {Meunier}, {Kral},
  {Henning}, {Menard}, {Moor}, {Avenhaus}, {Bazzon}, {Janson}, {Beuzit},
  {Bhowmik}, {Bonavita}, {Borgniet}, {Brandner}, {Cheetham}, {Cudel}, {Feldt},
  {Fontanive}, {Ginski}, {Hagelberg}, {Janin-Potiron}, {Lagadec}, {Langlois},
  {Le Coroller}, {Messina}, {Meyer}, {Mouillet}, {Peretti}, {Perrot}, {Rodet},
  {Samland}, {Sissa}, {Olofsson}, {Salter}, {Schmidt}, {Zurlo}, {Milli}, {van
  Boekel}, {Quanz}, {Feautrier}, {Le Mignant}, {Perret}, {Ramos}, \&
  {Rochat}}]{Chauvin:2018}
{Chauvin}, G., {Gratton}, R., {Bonnefoy}, M., {et~al.} 2018, \aap, 617, A76

\bibitem[{{Chein-I Chang}(2000)}]{Chang2000}
{Chein-I Chang}. 2000, IEEE Transactions on Information Theory, 46, 1927

\bibitem[{{Chein-I Chang}(2005)}]{Chang:2005}
{Chein-I Chang}. 2005, IEEE Transactions on Geoscience and Remote Sensing, 43,
  502

\bibitem[{{Chilcote} {et~al.}(2017){Chilcote}, {Pueyo}, {De Rosa}, {Vargas},
  {Macintosh}, {Bailey}, {Barman}, {Bauman}, {Bruzzone}, {Bulger}, {Burrows},
  {Cardwell}, {Chen}, {Cotten}, {Dillon}, {Doyon}, {Draper}, {Duch{\^e}ne},
  {Dunn}, {Erikson}, {Fitzgerald}, {Follette}, {Gavel}, {Goodsell}, {Graham},
  {Greenbaum}, {Hartung}, {Hibon}, {Hung}, {Ingraham}, {Kalas}, {Konopacky},
  {Larkin}, {Maire}, {Marchis}, {Marley}, {Marois}, {Metchev},
  {Millar-Blanchaer}, {Morzinski}, {Nielsen}, {Norton}, {Oppenheimer},
  {Palmer}, {Patience}, {Perrin}, {Poyneer}, {Rajan}, {Rameau},
  {Rantakyr{\"o}}, {Sadakuni}, {Saddlemyer}, {Savransky}, {Schneider}, {Serio},
  {Sivaramakrishnan}, {Song}, {Soummer}, {Thomas}, {Wallace}, {Wang},
  {Ward-Duong}, {Wiktorowicz}, \& {Wolff}}]{Chilcote:2017}
{Chilcote}, J., {Pueyo}, L., {De Rosa}, R.~J., {et~al.} 2017, \aj, 153, 182

\bibitem[{{Currie} {et~al.}(2018){Currie}, {Brandt}, {Uyama}, {Nielsen},
  {Blunt}, {Guyon}, {Tamura}, {Marois}, {Mede}, {Kuzuhara}, {Groff},
  {Jovanovic}, {Kasdin}, {Lozi}, {Hodapp}, {Chilcote}, {Carson}, {Martinache},
  {Goebel}, {Grady}, {McElwain}, {Akiyama}, {Asensio-Torres}, {Hayashi},
  {Janson}, {Knapp}, {Kwon}, {Nishikawa}, {Oh}, {Schlieder}, {Serabyn},
  {Sitko}, \& {Skaf}}]{Currie:2018}
{Currie}, T., {Brandt}, T.~D., {Uyama}, T., {et~al.} 2018, \aj, 156, 291

\bibitem[{{De Rosa} {et~al.}(2016){De Rosa}, {Rameau}, {Patience}, {Graham},
  {Doyon}, {Lafreni{\`e}re}, {Macintosh}, {Pueyo}, {Rajan}, {Wang},
  {Ward-Duong}, {Hung}, {Maire}, {Nielsen}, {Ammons}, {Bulger}, {Cardwell},
  {Chilcote}, {Galvez}, {Gerard}, {Goodsell}, {Hartung}, {Hibon}, {Ingraham},
  {Johnson-Groh}, {Kalas}, {Konopacky}, {Marchis}, {Marois}, {Metchev},
  {Morzinski}, {Oppenheimer}, {Perrin}, {Rantakyr{\"o}}, {Savransky}, \&
  {Thomas}}]{DeRosa:2016}
{De Rosa}, R.~J., {Rameau}, J., {Patience}, J., {et~al.} 2016, \apj, 824, 121

\bibitem[{{Delorme} {et~al.}(2017){Delorme}, {Schmidt}, {Bonnefoy}, {Desidera},
  {Ginski}, {Charnay}, {Lazzoni}, {Christiaens}, {Messina}, {D'Orazi}, {Milli},
  {Schlieder}, {Gratton}, {Rodet}, {Lagrange}, {Absil}, {Vigan}, {Galicher},
  {Hagelberg}, {Bonavita}, {Lavie}, {Zurlo}, {Olofsson}, {Boccaletti},
  {Cantalloube}, {Mouillet}, {Chauvin}, {Hambsch}, {Langlois}, {Udry},
  {Henning}, {Beuzit}, {Mordasini}, {Lucas}, {Marocco}, {Biller}, {Carson},
  {Cheetham}, {Covino}, {De Caprio}, {Delboulbe}, {Feldt}, {Girard}, {Hubin},
  {Maire}, {Pavlov}, {Petit}, {Rouan}, {Roelfsema}, \& {Wildi}}]{Delorme:2017}
{Delorme}, P., {Schmidt}, T., {Bonnefoy}, M., {et~al.} 2017, \aap, 608, A79

\bibitem[{{Deville} {et~al.}(2014){Deville}, {Revel}, {Achard}, \&
  {Briottet}}]{Deville:2014}
{Deville}, Y., {Revel}, C., {Achard}, V., \& {Briottet}, X. 2014, IEEE Whispers

\bibitem[{Drumetz(2016)}]{Drumetz:2016}
Drumetz, L. 2016, Theses, {Universit{\'e} Grenoble Alpes}

\bibitem[{Drumetz {et~al.}(2020)Drumetz, Chanussot, \& Jutten}]{Drumetz:2020}
Drumetz, L., Chanussot, J., \& Jutten, C. 2020, in Data Handling in Science and
  Technology, Vol.~32, Hyperspectral Imaging, ed. J.~M. Amigo (Elsevier), 167
  -- 203

\bibitem[{{Esposito} {et~al.}(2011){Esposito}, {Riccardi}, {Pinna}, {Puglisi},
  {Quir{\'o}s-Pacheco}, {Arcidiacono}, {Xompero}, {Briguglio}, {Agapito},
  {Busoni}, {Fini}, {Argomedo}, {Gherardi}, {Brusa}, {Miller}, {Guerra},
  {Stefanini}, \& {Salinari}}]{Esposito:2011}
{Esposito}, S., {Riccardi}, A., {Pinna}, E., {et~al.} 2011, in Society of
  Photo-Optical Instrumentation Engineers (SPIE) Conference Series, Vol. 8149,
  Astronomical Adaptive Optics Systems and Applications IV, ed. R.~K. {Tyson}
  \& M.~{Hart}, 814902

\bibitem[{{Flasseur} {et~al.}(2018){Flasseur}, {Denis}, {Thi{\'e}baut}, \&
  {Langlois}}]{Flasseur:2018}
{Flasseur}, O., {Denis}, L., {Thi{\'e}baut}, {\'E}., \& {Langlois}, M. 2018,
  \aap, 618, A138

\bibitem[{{Forni} {et~al.}(2005){Forni}, {Poulet}, {Bibring}, {Erard}, {Gomez},
  {Langevin}, {Gondet}, \& {OMEGA Science Team}}]{Forni:2005}
{Forni}, O., {Poulet}, F., {Bibring}, J.~P., {et~al.} 2005, in 36th Annual
  Lunar and Planetary Science Conference, ed. S.~{Mackwell} \& E.~{Stansbery},
  Lunar and Planetary Science Conference, 1623

\bibitem[{{Foschino} {et~al.}(2019){Foschino}, {Bern{\'e}}, \&
  {Joblin}}]{Foschino:2019}
{Foschino}, S., {Bern{\'e}}, O., \& {Joblin}, C. 2019, \aap, 632, A84

\bibitem[{{Galicher} {et~al.}(2018){Galicher}, {Boccaletti}, {Mesa}, {Delorme},
  {Gratton}, {Langlois}, {Lagrange}, {Maire}, {Le Coroller}, {Chauvin},
  {Biller}, {Cantalloube}, {Janson}, {Lagadec}, {Meunier}, {Vigan},
  {Hagelberg}, {Bonnefoy}, {Zurlo}, {Rocha}, {Maurel}, {Jaquet}, {Buey}, \&
  {Weber}}]{Galicher:2018}
{Galicher}, R., {Boccaletti}, A., {Mesa}, D., {et~al.} 2018, \aap, 615, A92

\bibitem[{{Gomez Gonzalez} {et~al.}(2016){Gomez Gonzalez}, {Absil}, {Absil},
  {Van Droogenbroeck}, {Mawet}, \& {Surdej}}]{Gomez:2016}
{Gomez Gonzalez}, C.~A., {Absil}, O., {Absil}, P.~A., {et~al.} 2016, \aap, 589,
  A54

\bibitem[{{Gratier} {et~al.}(2017){Gratier}, {Bron}, {Gerin}, {Pety}, {Guzman},
  {Orkisz}, {Bardeau}, {Goicoechea}, {Le Petit}, {Liszt}, {{\"O}berg},
  {Peretto}, {Roueff}, {Sievers}, \& {Tremblin}}]{Gratier:2017}
{Gratier}, P., {Bron}, E., {Gerin}, M., {et~al.} 2017, \aap, 599, A100

\bibitem[{{Greenbaum} {et~al.}(2018){Greenbaum}, {Pueyo}, {Ruffio}, {Wang}, {De
  Rosa}, {Aguilar}, {Rameau}, {Barman}, {Marois}, {Marley}, {Konopacky},
  {Rajan}, {Macintosh}, {Ansdell}, {Arriaga}, {Bailey}, {Bulger}, {Burrows},
  {Chilcote}, {Cotten}, {Doyon}, {Duch{\^e}ne}, {Fitzgerald}, {Follette},
  {Gerard}, {Goodsell}, {Graham}, {Hibon}, {Hung}, {Ingraham}, {Kalas},
  {Larkin}, {Maire}, {Marchis}, {Metchev}, {Millar-Blanchaer}, {Nielsen},
  {Norton}, {Oppenheimer}, {Palmer}, {Patience}, {Perrin}, {Poyneer},
  {Rantakyr{\"o}}, {Savransky}, {Schneider}, {Sivaramakrishnan}, {Song},
  {Soummer}, {Thomas}, {Wallace}, {Ward-Duong}, {Wiktorowicz}, \&
  {Wolff}}]{Greenbaum:2018}
{Greenbaum}, A.~Z., {Pueyo}, L., {Ruffio}, J.-B., {et~al.} 2018, \aj, 155, 226

\bibitem[{{Haffert} {et~al.}(2019){Haffert}, {Bohn}, {de Boer}, {Snellen},
  {Brinchmann}, {Girard}, {Keller}, \& {Bacon}}]{Haffert:2019}
{Haffert}, S.~Y., {Bohn}, A.~J., {de Boer}, J., {et~al.} 2019, Nature
  Astronomy, 3, 749

\bibitem[{Halimi {et~al.}(2015)Halimi, Honeine, Kharouf, Richard, \&
  Tourneret}]{Halimi:2015}
Halimi, A., Honeine, P., Kharouf, M., Richard, C., \& Tourneret, J.-Y. 2015,
  IEEE Transactions on Geoscience and Remote Sensing, 54

\bibitem[{{Harsanyi} \& {Chang}(1994)}]{Harsanyi:1994}
{Harsanyi}, J.~C. \& {Chang}, C.~. 1994, IEEE Transactions on Geoscience and
  Remote Sensing, 32, 779

\bibitem[{Hauksdottir {et~al.}(2006)Hauksdottir, Jutten, Schmidt, Chanussot,
  Benediktsson, \& Dout\'e}]{Hauksdottir06thephysical}
Hauksdottir, H., Jutten, C., Schmidt, F., {et~al.} 2006, in In 7th Nordic
  Signal Processing Symposium (NORSIG’2006

\bibitem[{{Hoeijmakers} {et~al.}(2018){Hoeijmakers}, {Schwarz}, {Snellen}, {de
  Kok}, {Bonnefoy}, {Chauvin}, {Lagrange}, \& {Girard}}]{Hoeijmakers:2018}
{Hoeijmakers}, H.~J., {Schwarz}, H., {Snellen}, I.~A.~G., {et~al.} 2018, \aap,
  617, A144

\bibitem[{{Houll{\'e}} {et~al.}(2021){Houll{\'e}}, {Vigan}, {Carlotti},
  {Choquet}, {Cantalloube}, {Phillips}, {Sauvage}, {Schwartz}, {Otten},
  {Baraffe}, {Emsenhuber}, \& {Mordasini}}]{Houlle:2021}
{Houll{\'e}}, M., {Vigan}, A., {Carlotti}, A., {et~al.} 2021, arXiv e-prints,
  arXiv:2104.11251

\bibitem[{{Jensen-Clem} {et~al.}(2018){Jensen-Clem}, {Mawet}, {Gomez Gonzalez},
  {Absil}, {Belikov}, {Currie}, {Kenworthy}, {Marois}, {Mazoyer}, {Ruane},
  {Tanner}, \& {Cantalloube}}]{Jensen:2018}
{Jensen-Clem}, R., {Mawet}, D., {Gomez Gonzalez}, C.~A., {et~al.} 2018, \aj,
  155, 19

\bibitem[{{Jovanovic} {et~al.}(2015){Jovanovic}, {Martinache}, {Guyon},
  {Clergeon}, {Singh}, {Kudo}, {Garrel}, {Newman}, {Doughty}, {Lozi}, {Males},
  {Minowa}, {Hayano}, {Takato}, {Morino}, {Kuhn}, {Serabyn}, {Norris},
  {Tuthill}, {Schworer}, {Stewart}, {Close}, {Huby}, {Perrin}, {Lacour},
  {Gauchet}, {Vievard}, {Murakami}, {Oshiyama}, {Baba}, {Matsuo}, {Nishikawa},
  {Tamura}, {Lai}, {Marchis}, {Duchene}, {Kotani}, \&
  {Woillez}}]{Jovanovic:2015}
{Jovanovic}, N., {Martinache}, F., {Guyon}, O., {et~al.} 2015, \pasp, 127, 890

\bibitem[{{Juvela} {et~al.}(1996){Juvela}, {Lehtinen}, \&
  {Paatero}}]{Juvela:1996}
{Juvela}, M., {Lehtinen}, K., \& {Paatero}, P. 1996, \mnras, 280, 616

\bibitem[{{Kawahara} {et~al.}(2014){Kawahara}, {Murakami}, {Matsuo}, \&
  {Kotani}}]{Kawahara:2014}
{Kawahara}, H., {Murakami}, N., {Matsuo}, T., \& {Kotani}, T. 2014, \apjs, 212,
  27

\bibitem[{{Keppler} {et~al.}(2018){Keppler}, {Benisty}, {M{\"u}ller},
  {Henning}, {van Boekel}, {Cantalloube}, {Ginski}, {van Holstein}, {Maire},
  {Pohl}, {Samland}, {Avenhaus}, {Baudino}, {Boccaletti}, {de Boer},
  {Bonnefoy}, {Chauvin}, {Desidera}, {Langlois}, {Lazzoni}, {Marleau},
  {Mordasini}, {Pawellek}, {Stolker}, {Vigan}, {Zurlo}, {Birnstiel},
  {Brandner}, {Feldt}, {Flock}, {Girard}, {Gratton}, {Hagelberg}, {Isella},
  {Janson}, {Juhasz}, {Kemmer}, {Kral}, {Lagrange}, {Launhardt}, {Matter},
  {M{\'e}nard}, {Milli}, {Molli{\`e}re}, {Olofsson}, {P{\'e}rez}, {Pinilla},
  {Pinte}, {Quanz}, {Schmidt}, {Udry}, {Wahhaj}, {Williams}, {Buenzli},
  {Cudel}, {Dominik}, {Galicher}, {Kasper}, {Lannier}, {Mesa}, {Mouillet},
  {Peretti}, {Perrot}, {Salter}, {Sissa}, {Wildi}, {Abe}, {Antichi},
  {Augereau}, {Baruffolo}, {Baudoz}, {Bazzon}, {Beuzit}, {Blanchard}, {Brems},
  {Buey}, {De Caprio}, {Carbillet}, {Carle}, {Cascone}, {Cheetham}, {Claudi},
  {Costille}, {Delboulb{\'e}}, {Dohlen}, {Fantinel}, {Feautrier}, {Fusco},
  {Giro}, {Gluck}, {Gry}, {Hubin}, {Hugot}, {Jaquet}, {Le Mignant}, {Llored},
  {Madec}, {Magnard}, {Martinez}, {Maurel}, {Meyer}, {M{\"o}ller-Nilsson},
  {Moulin}, {Mugnier}, {Orign{\'e}}, {Pavlov}, {Perret}, {Petit}, {Pragt},
  {Puget}, {Rabou}, {Ramos}, {Rigal}, {Rochat}, {Roelfsema}, {Rousset}, {Roux},
  {Salasnich}, {Sauvage}, {Sevin}, {Soenke}, {Stadler}, {Suarez}, {Turatto}, \&
  {Weber}}]{Keppler:2018}
{Keppler}, M., {Benisty}, M., {M{\"u}ller}, A., {et~al.} 2018, \aap, 617, A44

\bibitem[{{Konopacky} {et~al.}(2013){Konopacky}, {Barman}, {Macintosh}, \&
  {Marois}}]{Konopacky:2013}
{Konopacky}, Q.~M., {Barman}, T.~S., {Macintosh}, B.~A., \& {Marois}, C. 2013,
  Science, 339, 1398

\bibitem[{Kruse {et~al.}(1993)Kruse, Lefkoff, Boardman, Heidebrecht, Shapiro,
  Barloon, \& Goetz}]{Kruse:1993}
Kruse, F., Lefkoff, A., Boardman, J., {et~al.} 1993, Remote Sensing of
  Environment, 44, 145 , airbone Imaging Spectrometry

\bibitem[{{Kuntschner} {et~al.}(2014){Kuntschner}, {Jochum}, {Amico}, {Dekker},
  {Kerber}, {Marchetti}, {Accardo}, {Brast}, {Brinkmann}, {Conzelmann},
  {Delabre}, {Duchateau}, {Fedrigo}, {Finger}, {Frank}, {Rodriguez}, {Klein},
  {Knudstrup}, {Le Louarn}, {Lundin}, {Modigliani}, {M{\"u}ller}, {Neeser},
  {Tordo}, {Valenti}, {Eisenhauer}, {Sturm}, {Feuchtgruber}, {George}, {Hartl},
  {Hofmann}, {Huber}, {Plattner}, {Schubert}, {Tarantik}, {Wiezorrek}, {Meyer},
  {Quanz}, {Glauser}, {Weisz}, {Esposito}, {Xompero}, {Agapito}, {Antichi},
  {Biliotti}, {Bonaglia}, {Briguglio}, {Carbonaro}, {Cresci}, {Fini}, {Pinna},
  {Puglisi}, {Quir{\'o}s-Pacheco}, {Riccardi}, {Di Rico}, {Arcidiacono}, \&
  {Dolci}}]{Kuntscner:2014}
{Kuntschner}, H., {Jochum}, L., {Amico}, P., {et~al.} 2014, in Society of
  Photo-Optical Instrumentation Engineers (SPIE) Conference Series, Vol. 9147,
  Ground-based and Airborne Instrumentation for Astronomy V, ed. S.~K.
  {Ramsay}, I.~S. {McLean}, \& H.~{Takami}, 91471U

\bibitem[{{Lafreni{\`e}re} {et~al.}(2007){Lafreni{\`e}re}, {Marois}, {Doyon},
  {Nadeau}, \& {Artigau}}]{Lafreniere:2007}
{Lafreni{\`e}re}, D., {Marois}, C., {Doyon}, R., {Nadeau}, D., \& {Artigau},
  {\'E}. 2007, \apj, 660, 770

\bibitem[{{Lagrange} {et~al.}(2019){Lagrange}, {Boccaletti}, {Langlois},
  {Chauvin}, {Gratton}, {Beust}, {Desidera}, {Milli}, {Bonnefoy}, {Cheetham},
  {Feldt}, {Meyer}, {Vigan}, {Biller}, {Bonavita}, {Baudino}, {Cantalloube},
  {Cudel}, {Daemgen}, {Delorme}, {D'Orazi}, {Girard}, {Fontanive}, {Hagelberg},
  {Janson}, {Keppler}, {Koypitova}, {Galicher}, {Lannier}, {Le Coroller},
  {Ligi}, {Maire}, {Mesa}, {Messina}, {M{\"u}eller}, {Peretti}, {Perrot},
  {Rouan}, {Salter}, {Samland}, {Schmidt}, {Sissa}, {Zurlo}, {Beuzit},
  {Mouillet}, {Dominik}, {Henning}, {Lagadec}, {M{\'e}nard}, {Schmid},
  {Turatto}, {Udry}, {Bohn}, {Charnay}, {Gomez Gonzales}, {Gry}, {Kenworthy},
  {Kral}, {Mordasini}, {Moutou}, {van der Plas}, {Schlieder}, {Abe}, {Antichi},
  {Baruffolo}, {Baudoz}, {Baudrand}, {Blanchard}, {Bazzon}, {Buey},
  {Carbillet}, {Carle}, {Charton}, {Cascone}, {Claudi}, {Costille}, {Deboulbe},
  {De Caprio}, {Dohlen}, {Fantinel}, {Feautrier}, {Fusco}, {Gigan}, {Giro},
  {Gisler}, {Gluck}, {Hubin}, {Hugot}, {Jaquet}, {Kasper}, {Madec}, {Magnard},
  {Martinez}, {Maurel}, {Le Mignant}, {M{\"o}ller-Nilsson}, {Llored}, {Moulin},
  {Orign{\'e}}, {Pavlov}, {Perret}, {Petit}, {Pragt}, {Szulagyi}, \&
  {Wildi}}]{2019A&A...621L...8L}
{Lagrange}, A.~M., {Boccaletti}, A., {Langlois}, M., {et~al.} 2019, \aap, 621,
  L8

\bibitem[{{Lagrange} {et~al.}(2010){Lagrange}, {Bonnefoy}, {Chauvin}, {Apai},
  {Ehrenreich}, {Boccaletti}, {Gratadour}, {Rouan}, {Mouillet}, {Lacour}, \&
  {Kasper}}]{Lagrange:2010}
{Lagrange}, A.~M., {Bonnefoy}, M., {Chauvin}, G., {et~al.} 2010, Science, 329,
  57

\bibitem[{{Larkin} {et~al.}(2016){Larkin}, {Moore}, {Wright}, {Wincentsen},
  {Anderson}, {Chisholm}, {Dekany}, {Dunn}, {Ellerbroek}, {Hayano}, {Phillips},
  {Simard}, {Smith}, {Suzuki}, {Weber}, {Weiss}, \& {Zhang}}]{Larkin:2016}
{Larkin}, J.~E., {Moore}, A.~M., {Wright}, S.~A., {et~al.} 2016, in Society of
  Photo-Optical Instrumentation Engineers (SPIE) Conference Series, Vol. 9908,
  Ground-based and Airborne Instrumentation for Astronomy VI, ed. C.~J.
  {Evans}, L.~{Simard}, \& H.~{Takami}, 99081W

\bibitem[{{Li} {et~al.}(2015){Li}, {Zhang}, {Zhang}, \& {Ma}}]{Li:2015}
{Li}, J., {Zhang}, H., {Zhang}, L., \& {Ma}, L. 2015, IEEE Journal of Selected
  Topics in Applied Earth Observations and Remote Sensing, 8, 2523

\bibitem[{Liu {et~al.}(2018)Liu, Luo, Dout\'e, \& Chanussot}]{Liu:2018}
Liu, J., Luo, B., Dout\'e, S., \& Chanussot, J. 2018, Remote Sensing, 10, 737

\bibitem[{{Lovis} {et~al.}(2017){Lovis}, {Snellen}, {Mouillet}, {Pepe},
  {Wildi}, {Astudillo-Defru}, {Beuzit}, {Bonfils}, {Cheetham}, {Conod},
  {Delfosse}, {Ehrenreich}, {Figueira}, {Forveille}, {Martins}, {Quanz},
  {Santos}, {Schmid}, {S{\'e}gransan}, \& {Udry}}]{Lovis:2017}
{Lovis}, C., {Snellen}, I., {Mouillet}, D., {et~al.} 2017, \aap, 599, A16

\bibitem[{Luo {et~al.}(2013)Luo, Chanussot, Dout\'e, \& Zhang}]{Luo:2013}
Luo, B., Chanussot, J., Dout\'e, S., \& Zhang, L. 2013, IEEE Geoscience and
  Remote Sensing Letters, 10, 24

\bibitem[{{Macintosh} {et~al.}(2015){Macintosh}, {Graham}, {Barman}, {De Rosa},
  {Konopacky}, {Marley}, {Marois}, {Nielsen}, {Pueyo}, {Rajan}, {Rameau},
  {Saumon}, {Wang}, {Patience}, {Ammons}, {Arriaga}, {Artigau}, {Beckwith},
  {Brewster}, {Bruzzone}, {Bulger}, {Burningham}, {Burrows}, {Chen}, {Chiang},
  {Chilcote}, {Dawson}, {Dong}, {Doyon}, {Draper}, {Duch{\^e}ne}, {Esposito},
  {Fabrycky}, {Fitzgerald}, {Follette}, {Fortney}, {Gerard}, {Goodsell},
  {Greenbaum}, {Hibon}, {Hinkley}, {Cotten}, {Hung}, {Ingraham},
  {Johnson-Groh}, {Kalas}, {Lafreniere}, {Larkin}, {Lee}, {Line}, {Long},
  {Maire}, {Marchis}, {Matthews}, {Max}, {Metchev}, {Millar-Blanchaer},
  {Mittal}, {Morley}, {Morzinski}, {Murray-Clay}, {Oppenheimer}, {Palmer},
  {Patel}, {Perrin}, {Poyneer}, {Rafikov}, {Rantakyr{\"o}}, {Rice}, {Rojo},
  {Rudy}, {Ruffio}, {Ruiz}, {Sadakuni}, {Saddlemyer}, {Salama}, {Savransky},
  {Schneider}, {Sivaramakrishnan}, {Song}, {Soummer}, {Thomas}, {Vasisht},
  {Wallace}, {Ward-Duong}, {Wiktorowicz}, {Wolff}, \&
  {Zuckerman}}]{Macintosh:2015}
{Macintosh}, B., {Graham}, J.~R., {Barman}, T., {et~al.} 2015, Science, 350, 64

\bibitem[{{Macintosh} {et~al.}(2014){Macintosh}, {Graham}, {Ingraham},
  {Konopacky}, {Marois}, {Perrin}, {Poyneer}, {Bauman}, {Barman}, {Burrows},
  {Cardwell}, {Chilcote}, {De Rosa}, {Dillon}, {Doyon}, {Dunn}, {Erikson},
  {Fitzgerald}, {Gavel}, {Goodsell}, {Hartung}, {Hibon}, {Kalas}, {Larkin},
  {Maire}, {Marchis}, {Marley}, {McBride}, {Millar-Blanchaer}, {Morzinski},
  {Norton}, {Oppenheimer}, {Palmer}, {Patience}, {Pueyo}, {Rantakyro},
  {Sadakuni}, {Saddlemyer}, {Savransky}, {Serio}, {Soummer},
  {Sivaramakrishnan}, {Song}, {Thomas}, {Wallace}, {Wiktorowicz}, \&
  {Wolff}}]{Macintosh:2014}
{Macintosh}, B., {Graham}, J.~R., {Ingraham}, P., {et~al.} 2014, Proceedings of
  the National Academy of Science, 111, 12661

\bibitem[{{Marois} {et~al.}(2003){Marois}, {Doyon}, {Nadeau}, {Racine},
  {Riopel}, \& {Vallee}}]{Marois:2003}
{Marois}, C., {Doyon}, R., {Nadeau}, D., {et~al.} 2003, in Society of
  Photo-Optical Instrumentation Engineers (SPIE) Conference Series, Vol. 4860,
  High-Contrast Imaging for Exo-Planet Detection., ed. A.~B. {Schultz},
  130--137

\bibitem[{{Marois} {et~al.}(2006){Marois}, {Lafreni{\`e}re}, {Doyon},
  {Macintosh}, \& {Nadeau}}]{Marois:2006}
{Marois}, C., {Lafreni{\`e}re}, D., {Doyon}, R., {Macintosh}, B., \& {Nadeau},
  D. 2006, \apj, 641, 556

\bibitem[{{Mawet} {et~al.}(2018){Mawet}, {Bond}, {Delorme}, {Jovanovic},
  {Cetre}, {Chun}, {Echeverri}, {Hall}, {Lilley}, {Wallace}, \&
  {Wizinowich}}]{Mawet:2018}
{Mawet}, D., {Bond}, C.~Z., {Delorme}, J.~R., {et~al.} 2018, in Society of
  Photo-Optical Instrumentation Engineers (SPIE) Conference Series, Vol. 10703,
  Adaptive Optics Systems VI, ed. L.~M. {Close}, L.~{Schreiber}, \&
  D.~{Schmidt}, 1070306

\bibitem[{{Mawet} {et~al.}(2014){Mawet}, {Milli}, {Wahhaj}, {Pelat}, {Absil},
  {Delacroix}, {Boccaletti}, {Kasper}, {Kenworthy}, {Marois}, {Mennesson}, \&
  {Pueyo}}]{Mawet:2014}
{Mawet}, D., {Milli}, J., {Wahhaj}, Z., {et~al.} 2014, \apj, 792, 97

\bibitem[{{McGregor} {et~al.}(2012){McGregor}, {Bloxham}, {Boz}, {Davies},
  {Doolan}, {Ellis}, {Hart}, {Jones}, {Luvaul}, {Nielsen}, {Parcell}, {Sharp},
  {Stevanovic}, \& {Young}}]{McGregor:2012}
{McGregor}, P.~J., {Bloxham}, G.~J., {Boz}, R., {et~al.} 2012, in Society of
  Photo-Optical Instrumentation Engineers (SPIE) Conference Series, Vol. 8446,
  Ground-based and Airborne Instrumentation for Astronomy IV, ed. I.~S.
  {McLean}, S.~K. {Ramsay}, \& H.~{Takami}, 84461I

\bibitem[{{Mesa} {et~al.}(2015){Mesa}, {Gratton}, {Zurlo}, {Vigan}, {Claudi},
  {Alberi}, {Antichi}, {Baruffolo}, {Beuzit}, {Boccaletti}, {Bonnefoy},
  {Costille}, {Desidera}, {Dohlen}, {Fantinel}, {Feldt}, {Fusco}, {Giro},
  {Henning}, {Kasper}, {Langlois}, {Maire}, {Martinez}, {Moeller-Nilsson},
  {Mouillet}, {Moutou}, {Pavlov}, {Puget}, {Salasnich}, {Sauvage}, {Sissa},
  {Turatto}, {Udry}, {Vakili}, {Waters}, \& {Wildi}}]{Mesa:2015}
{Mesa}, D., {Gratton}, R., {Zurlo}, A., {et~al.} 2015, \aap, 576, A121

\bibitem[{Moussaoui {et~al.}(2008)Moussaoui, Hauksdottir, Schmidt, Jutten,
  Chanussot, Brie, Dout\'e, \& Benediktsson}]{MOUSSAOUI20082194}
Moussaoui, S., Hauksdottir, H., Schmidt, F., {et~al.} 2008, Neurocomputing, 71,
  2194 , neurocomputing for Vision Research Advances in Blind Signal Processing

\bibitem[{{M{\"u}ller} {et~al.}(2018){M{\"u}ller}, {Keppler}, {Henning},
  {Samland}, {Chauvin}, {Beust}, {Maire}, {Molaverdikhani}, {van Boekel},
  {Benisty}, {Boccaletti}, {Bonnefoy}, {Cantalloube}, {Charnay}, {Baudino},
  {Gennaro}, {Long}, {Cheetham}, {Desidera}, {Feldt}, {Fusco}, {Girard},
  {Gratton}, {Hagelberg}, {Janson}, {Lagrange}, {Langlois}, {Lazzoni}, {Ligi},
  {M{\'e}nard}, {Mesa}, {Meyer}, {Molli{\`e}re}, {Mordasini}, {Moulin},
  {Pavlov}, {Pawellek}, {Quanz}, {Ramos}, {Rouan}, {Sissa}, {Stadler}, {Vigan},
  {Wahhaj}, {Weber}, \& {Zurlo}}]{Muller:2018}
{M{\"u}ller}, A., {Keppler}, M., {Henning}, T., {et~al.} 2018, \aap, 617, L2

\bibitem[{{Nielsen} {et~al.}(2019){Nielsen}, {De Rosa}, {Macintosh}, {Wang},
  {Ruffio}, {Chiang}, {Marley}, {Saumon}, {Savransky}, {Fabrycky}, {Konopacky},
  {Patience}, \& {Bailey}}]{Nielsen:2019}
{Nielsen}, E., {De Rosa}, R., {Macintosh}, B., {et~al.} 2019, in AAS/Division
  for Extreme Solar Systems Abstracts, Vol.~51, AAS/Division for Extreme Solar
  Systems Abstracts, 100.02

\bibitem[{Nowak {et~al.}(2020)Nowak, Lacour, Molli\'ere, Wang, Charnay,
  Dishoeck, Abuter, Amorim, Berger, Beust, Bonnefoy, Bonnet, Brandner, Buron,
  Cantalloube, Collin, Chapron, Cl\'enet, Foresto, Zeeuw, Dembet, Dexter,
  Duvert, Eckart, Eisenhauer, Schreiber, F\'edou, Lopez, Gao, Gendron, Genzel,
  Gillessen, Haussmann, Henning, Hippler, Hubert, Jocou, Kervella, Lagrange,
  Lapeyr\`ere, Bouquin, Lana, Maire, Ott, Paumard, Paladini, Perraut, Perrin,
  Pueyo, Pfuhl, Rabien, Rau, Rodriguez-Coira, Rousset, Scheithauer, Shangguan,
  Straub, Straubmeier, Sturm, Tacconi, Vincent, Widmann, Wieprecht, Wiezorrek,
  Woillez, Yazici, \& Ziegler}]{Nowak:2020}
Nowak, M., Lacour, S., Molli\'ere, P., {et~al.} 2020, Astronomy \&
  Astrophysics, 633, A110, publisher: EDP Sciences

\bibitem[{{Otten} {et~al.}(2021){Otten}, {Vigan}, {Muslimov}, {N'Diaye},
  {Choquet}, {Seemann}, {Dohlen}, {Houll{\'e}}, {Cristofari}, {Phillips},
  {Charles}, {Baraffe}, {Beuzit}, {Costille}, {Dorn}, {El Morsy}, {Kasper},
  {Lopez}, {Mordasini}, {Pourcelot}, {Reiners}, \& {Sauvage}}]{Otten:2021}
{Otten}, G.~P.~P.~L., {Vigan}, A., {Muslimov}, E., {et~al.} 2021, \aap, 646,
  A150

\bibitem[{{Petit dit de la Roche} {et~al.}(2018){Petit dit de la Roche},
  {Hoeijmakers}, \& {Snellen}}]{PetitdelaRoche:2018}
{Petit dit de la Roche}, D.~J.~M., {Hoeijmakers}, H.~J., \& {Snellen}, I.~A.~G.
  2018, \aap, 616, A146

\bibitem[{{Petrus} {et~al.}(2020){Petrus}, {Bonnefoy}, {Chauvin}, {Charnay},
  {Marleau}, {Gratton}, {Lagrange}, {Rameau}, {Mordasini}, {Nowak}, {Delorme},
  {Boccaletti}, {Carlotti}, {Houll{\'e}}, {Vigan}, {Allard}, {Desidera},
  {D'Orazi}, {Hoeijmakers}, {Wyttenbach}, \& {Lavie}}]{Petrus:2020}
{Petrus}, S., {Bonnefoy}, M., {Chauvin}, G., {et~al.} 2020, arXiv e-prints,
  arXiv:2012.02798

\bibitem[{{Plaza} {et~al.}(2004){Plaza}, {Martinez}, {Perez}, \&
  {Plaza}}]{Plaza:2006}
{Plaza}, A., {Martinez}, P., {Perez}, R., \& {Plaza}, J. 2004, IEEE
  Transactions on Geoscience and Remote Sensing, 42, 650

\bibitem[{{Racine} {et~al.}(1999){Racine}, {Walker}, {Nadeau}, {Doyon}, \&
  {Marois}}]{Racine:1999}
{Racine}, R., {Walker}, G. A.~H., {Nadeau}, D., {Doyon}, R., \& {Marois}, C.
  1999, \pasp, 111, 587

\bibitem[{{Rajan} {et~al.}(2017){Rajan}, {Rameau}, {De Rosa}, {Marley},
  {Graham}, {Macintosh}, {Marois}, {Morley}, {Patience}, {Pueyo}, {Saumon},
  {Ward-Duong}, {Ammons}, {Arriaga}, {Bailey}, {Barman}, {Bulger}, {Burrows},
  {Chilcote}, {Cotten}, {Czekala}, {Doyon}, {Duch{\^e}ne}, {Esposito},
  {Fitzgerald}, {Follette}, {Fortney}, {Goodsell}, {Greenbaum}, {Hibon},
  {Hung}, {Ingraham}, {Johnson-Groh}, {Kalas}, {Konopacky}, {Lafreni{\`e}re},
  {Larkin}, {Maire}, {Marchis}, {Metchev}, {Millar-Blanchaer}, {Morzinski},
  {Nielsen}, {Oppenheimer}, {Palmer}, {Patel}, {Perrin}, {Poyneer},
  {Rantakyr{\"o}}, {Ruffio}, {Savransky}, {Schneider}, {Sivaramakrishnan},
  {Song}, {Soummer}, {Thomas}, {Vasisht}, {Wallace}, {Wang}, {Wiktorowicz}, \&
  {Wolff}}]{Rajan:2017}
{Rajan}, A., {Rameau}, J., {De Rosa}, R.~J., {et~al.} 2017, \aj, 154, 10

\bibitem[{{Rapacioli} {et~al.}(2005){Rapacioli}, {Joblin}, \&
  {Boissel}}]{Rapacioli:2005}
{Rapacioli}, M., {Joblin}, C., \& {Boissel}, P. 2005, \aap, 429, 193

\bibitem[{{Ren} {et~al.}(2018){Ren}, {Pueyo}, {Zhu}, {Debes}, \&
  {Duch{\^e}ne}}]{Ren:2018}
{Ren}, B., {Pueyo}, L., {Zhu}, G.~B., {Debes}, J., \& {Duch{\^e}ne}, G. 2018,
  \apj, 852, 104

\bibitem[{{Riaud} \& {Schneider}(2007)}]{Riaud:2007}
{Riaud}, P. \& {Schneider}, J. 2007, \aap, 469, 355

\bibitem[{Rieke {et~al.}(2015)Rieke, Wright, Buker, Bouwman, Colina, Glasse,
  Gordon, Greene, Godel, Henning, Justtanont, Lagage, Meixner,
  N{\o}rgaard-Nielsen, Ray, Ressler, van Dishoeck, \& Waelkens}]{Rieke:2015}
Rieke, G.~H., Wright, G.~S., Buker, T., {et~al.} 2015, Publications of the
  Astronomical Society of the Pacific, 127, 584

\bibitem[{{Ruffio} {et~al.}(2019){Ruffio}, {Macintosh}, {Konopacky}, {Barman},
  {De Rosa}, {Wang}, {Wilcomb}, {Czekala}, \& {Marois}}]{Ruffio:2019}
{Ruffio}, J.-B., {Macintosh}, B., {Konopacky}, Q.~M., {et~al.} 2019, \aj, 158,
  200

\bibitem[{{Ruffio} {et~al.}(2017){Ruffio}, {Macintosh}, {Wang}, {Pueyo},
  {Nielsen}, {De Rosa}, {Czekala}, {Marley}, {Arriaga}, {Bailey}, {Barman},
  {Bulger}, {Chilcote}, {Cotten}, {Doyon}, {Duch{\^e}ne}, {Fitzgerald},
  {Follette}, {Gerard}, {Goodsell}, {Graham}, {Greenbaum}, {Hibon}, {Hung},
  {Ingraham}, {Kalas}, {Konopacky}, {Larkin}, {Maire}, {Marchis}, {Marois},
  {Metchev}, {Millar-Blanchaer}, {Morzinski}, {Oppenheimer}, {Palmer},
  {Patience}, {Perrin}, {Poyneer}, {Rajan}, {Rameau}, {Rantakyr{\"o}},
  {Savransky}, {Schneider}, {Sivaramakrishnan}, {Song}, {Soummer}, {Thomas},
  {Wallace}, {Ward-Duong}, {Wiktorowicz}, \& {Wolff}}]{Ruffio:2017}
{Ruffio}, J.-B., {Macintosh}, B., {Wang}, J.~J., {et~al.} 2017, \apj, 842, 14

\bibitem[{{Samland} {et~al.}(2020){Samland}, {Bouwman}, {Hogg}, {Brandner},
  {Henning}, \& {Janson}}]{Samland:2020}
{Samland}, M., {Bouwman}, J., {Hogg}, D.~W., {et~al.} 2020, arXiv e-prints,
  arXiv:2011.12311

\bibitem[{{Samland} {et~al.}(2017){Samland}, {Molli{\`e}re}, {Bonnefoy},
  {Maire}, {Cantalloube}, {Cheetham}, {Mesa}, {Gratton}, {Biller}, {Wahhaj},
  {Bouwman}, {Brandner}, {Melnick}, {Carson}, {Janson}, {Henning}, {Homeier},
  {Mordasini}, {Langlois}, {Quanz}, {van Boekel}, {Zurlo}, {Schlieder},
  {Avenhaus}, {Beuzit}, {Boccaletti}, {Bonavita}, {Chauvin}, {Claudi}, {Cudel},
  {Desidera}, {Feldt}, {Fusco}, {Galicher}, {Kopytova}, {Lagrange}, {Le
  Coroller}, {Martinez}, {Moeller-Nilsson}, {Mouillet}, {Mugnier}, {Perrot},
  {Sevin}, {Sissa}, {Vigan}, \& {Weber}}]{Samland:2017}
{Samland}, M., {Molli{\`e}re}, P., {Bonnefoy}, M., {et~al.} 2017, \aap, 603,
  A57

\bibitem[{{Schwartz} {et~al.}(2016){Schwartz}, {Sekowski}, {Haggard},
  {Pall{\'e}}, \& {Cowan}}]{Schwartz:2016}
{Schwartz}, J.~C., {Sekowski}, C., {Haggard}, H.~M., {Pall{\'e}}, E., \&
  {Cowan}, N.~B. 2016, \mnras, 457, 926

\bibitem[{{Skemer} {et~al.}(2015){Skemer}, {Hinz}, {Montoya}, {Skrutskie},
  {Leisenring}, {Durney}, {Woodward}, {Wilson}, {Nelson}, {Bailey}, {Defrere},
  \& {Stone}}]{Skemer:2015}
{Skemer}, A.~J., {Hinz}, P., {Montoya}, M., {et~al.} 2015, in Society of
  Photo-Optical Instrumentation Engineers (SPIE) Conference Series, Vol. 9605,
  Techniques and Instrumentation for Detection of Exoplanets VII, ed.
  S.~{Shaklan}, 96051D

\bibitem[{{Snellen} {et~al.}(2015){Snellen}, {de Kok}, {Birkby}, {Brandl},
  {Brogi}, {Keller}, {Kenworthy}, {Schwarz}, \& {Stuik}}]{Snellen:2015}
{Snellen}, I., {de Kok}, R., {Birkby}, J.~L., {et~al.} 2015, \aap, 576, A59

\bibitem[{{Snellen} {et~al.}(2014){Snellen}, {Brandl}, {de Kok}, {Brogi},
  {Birkby}, \& {Schwarz}}]{Snellen:2014}
{Snellen}, I. A.~G., {Brandl}, B.~R., {de Kok}, R.~J., {et~al.} 2014, \nat,
  509, 63

\bibitem[{Soofbaf {et~al.}(2018)Soofbaf, Sahebi, \& Mojaradi}]{Soofbaf:2018}
Soofbaf, S.~R., Sahebi, M., \& Mojaradi, B. 2018, Remote Sensing, 10, 434

\bibitem[{{Soummer} {et~al.}(2012){Soummer}, {Pueyo}, \&
  {Larkin}}]{Soummer:2012}
{Soummer}, R., {Pueyo}, L., \& {Larkin}, J. 2012, \apjl, 755, L28

\bibitem[{{Sparks} \& {Ford}(2002)}]{Sparks:2002}
{Sparks}, W.~B. \& {Ford}, H.~C. 2002, \apj, 578, 543

\bibitem[{{Stone} {et~al.}(2020){Stone}, {Barman}, {Skemer}, {Briesemeister},
  {Brock}, {Hinz}, {Leisenring}, {Woodward}, {Skrutskie}, \&
  {Spalding}}]{Stone:2020}
{Stone}, J.~M., {Barman}, T., {Skemer}, A.~J., {et~al.} 2020, \aj, 160, 262

\bibitem[{{Thatte} {et~al.}(2016){Thatte}, {Clarke}, {Bryson}, {Shnetler},
  {Tecza}, {Fusco}, {Bacon}, {Richard}, {Mediavilla}, {Neichel}, {Arribas},
  {Garcia-Lorenzo}, {Evans}, {Remillieux}, {El Madi}, {Herreros}, {Melotte},
  {O'Brien}, {Tosh}, {Vernet}, {Hammersley}, {Ives}, {Finger}, {Houghton},
  {Rigopoulou}, {Lynn}, {Allen}, {Zieleniewski}, {Kendrew}, {Ferraro-Wood},
  {P{\'e}contal-Rousset}, {Kosmalski}, {Laurent}, {Loupias}, {Piqueras},
  {Renault}, {Blaizot}, {Daguis{\'e}}, {Migniau}, {Jarno}, {Born}, {Gallie},
  {Montgomery}, {Henry}, {Schwartz}, {Taylor}, {Zins}, {Rodr{\'\i}guez-Ramos},
  {Cagigas}, {Battaglia}, {Rebolo L{\'o}pez}, {Hern{\'a}ndez Su{\'a}rez},
  {Gigante-Ripoll}, {Piqueras L{\'o}pez}, {Villa Martin}, {Correia}, {Pascal},
  {Blanco}, {Vola}, {Epinat}, {Peroux}, {Vigan}, {Dohlen}, {Sauvage}, {Lee},
  {Carlotti}, {Verinaud}, {Morris}, {Myers}, {Reeves}, {Swinbank}, {Calcines},
  \& {Larrieu}}]{Thatte:2016}
{Thatte}, N.~A., {Clarke}, F., {Bryson}, I., {et~al.} 2016, in Society of
  Photo-Optical Instrumentation Engineers (SPIE) Conference Series, Vol. 9908,
  Ground-based and Airborne Instrumentation for Astronomy VI, ed. C.~J.
  {Evans}, L.~{Simard}, \& H.~{Takami}, 99081X

\bibitem[{{Themelis} {et~al.}(2012){Themelis}, {Schmidt}, {Sykioti},
  {Rontogiannis}, {Koutroumbas}, \& {Daglis}}]{themelis:2011}
{Themelis}, K.~E., {Schmidt}, F., {Sykioti}, O., {et~al.} 2012, \planss, 68, 34

\bibitem[{{Tsinos} {et~al.}(2017){Tsinos}, {Rontogiannis}, \&
  {Berberidis}}]{Tsinos:2017}
{Tsinos}, C.~G., {Rontogiannis}, A.~A., \& {Berberidis}, K. 2017, IEEE
  Transactions on Computational Imaging, 3, 160

\bibitem[{{Uyama} {et~al.}(2020){Uyama}, {Currie}, {Hori}, {De Rosa}, {Mede},
  {Brandt}, {Kwon}, {Guyon}, {Lozi}, {Jovanovic}, {Martinache}, {Kudo},
  {Tamura}, {Kasdin}, {Groff}, {Chilcote}, {Hayashi}, {McElwain},
  {Asensio-Torres}, {Janson}, {Knapp}, \& {Serabyn}}]{Uyama:2020}
{Uyama}, T., {Currie}, T., {Hori}, Y., {et~al.} 2020, \aj, 159, 40

\bibitem[{{Vigan} {et~al.}(2020){Vigan}, {Fontanive}, {Meyer}, {Biller},
  {Bonavita}, {Feldt}, {Desidera}, {Marleau}, {Emsenhuber}, {Galicher}, {Rice},
  {Forgan}, {Mordasini}, {Gratton}, {Le Coroller}, {Maire}, {Cantalloube},
  {Chauvin}, {Cheetham}, {Hagelberg}, {Lagrange}, {Langlois}, {Bonnefoy},
  {Beuzit}, {Boccaletti}, {D'Orazi}, {Delorme}, {Dominik}, {Henning}, {Janson},
  {Lagadec}, {Lazzoni}, {Ligi}, {Menard}, {Mesa}, {Messina}, {Moutou},
  {M{\"u}ller}, {Perrot}, {Samland}, {Schmid}, {Schmidt}, {Sissa}, {Turatto},
  {Udry}, {Zurlo}, {Abe}, {Antichi}, {Asensio-Torres}, {Baruffolo}, {Baudoz},
  {Baudrand}, {Bazzon}, {Blanchard}, {Bohn}, {Brown Sevilla}, {Carbillet},
  {Carle}, {Cascone}, {Charton}, {Claudi}, {Costille}, {De Caprio},
  {Delboulb{\'e}}, {Dohlen}, {Engler}, {Fantinel}, {Feautrier}, {Fusco},
  {Gigan}, {Girard}, {Giro}, {Gisler}, {Gluck}, {Gry}, {Hubin}, {Hugot},
  {Jaquet}, {Kasper}, {Le Mignant}, {Llored}, {Madec}, {Magnard}, {Martinez},
  {Maurel}, {M{\"o}ller-Nilsson}, {Mouillet}, {Moulin}, {Orign{\'e}}, {Pavlov},
  {Perret}, {Petit}, {Pragt}, {Puget}, {Rabou}, {Ramos}, {Rickman}, {Rigal},
  {Rochat}, {Roelfsema}, {Rousset}, {Roux}, {Salasnich}, {Sauvage}, {Sevin},
  {Soenke}, {Stadler}, {Suarez}, {Wahhaj}, {Weber}, \& {Wildi}}]{Vigan:2020}
{Vigan}, A., {Fontanive}, C., {Meyer}, M., {et~al.} 2020, arXiv e-prints,
  arXiv:2007.06573

\bibitem[{{Wang} {et~al.}(2018){Wang}, {Mawet}, {Fortney}, {Hood}, {Morley}, \&
  {Benneke}}]{Wang:2018}
{Wang}, J., {Mawet}, D., {Fortney}, J.~J., {et~al.} 2018, \aj, 156, 272

\bibitem[{{Wang} {et~al.}(2017){Wang}, {Mawet}, {Ruane}, {Hu}, \&
  {Benneke}}]{Wang:2017}
{Wang}, J., {Mawet}, D., {Ruane}, G., {Hu}, R., \& {Benneke}, B. 2017, \aj,
  153, 183

\bibitem[{{Wang} {et~al.}(2015){Wang}, {Ruffio}, {De Rosa}, {Aguilar}, {Wolff},
  \& {Pueyo}}]{WangJ:2015}
{Wang}, J.~J., {Ruffio}, J.-B., {De Rosa}, R.~J., {et~al.} 2015, {pyKLIP: PSF
  Subtraction for Exoplanets and Disks}

\bibitem[{{Ward-Duong} {et~al.}(2021){Ward-Duong}, {Patience}, {Follette}, {De
  Rosa}, {Rameau}, {Marley}, {Saumon}, {Nielsen}, {Rajan}, {Greenbaum}, {Lee},
  {Wang}, {Czekala}, {Duch{\^e}ne}, {Macintosh}, {Ammons}, {Bailey}, {Barman},
  {Bulger}, {Chen}, {Chilcote}, {Cotten}, {Doyon}, {Esposito}, {Fitzgerald},
  {Gerard}, {Goodsell}, {Graham}, {Hibon}, {Hom}, {Hung}, {Ingraham}, {Kalas},
  {Konopacky}, {Larkin}, {Maire}, {Marchis}, {Marois}, {Metchev},
  {Millar-Blanchaer}, {Oppenheimer}, {Palmer}, {Perrin}, {Poyneer}, {Pueyo},
  {Rantakyr{\"o}}, {Ren}, {Ruffio}, {Savransky}, {Schneider},
  {Sivaramakrishnan}, {Song}, {Soummer}, {Tallis}, {Thomas}, {Wallace},
  {Wiktorowicz}, \& {Wolff}}]{Ward:2021}
{Ward-Duong}, K., {Patience}, J., {Follette}, K., {et~al.} 2021, \aj, 161, 5

\bibitem[{{Wilcomb} {et~al.}(2020){Wilcomb}, {Konopacky}, {Barman}, {Theissen},
  {Ruffio}, {Brock}, {Macintosh}, \& {Marois}}]{Wilcomb:2020}
{Wilcomb}, K.~K., {Konopacky}, Q.~M., {Barman}, T.~S., {et~al.} 2020, \aj, 160,
  207

\bibitem[{{Xie} {et~al.}(2020){Xie}, {Haffert}, {de Boer}, {Kenworthy},
  {Brinchmann}, {Girard}, {Snellen}, \& {Keller}}]{Xie:2020}
{Xie}, C., {Haffert}, S.~Y., {de Boer}, J., {et~al.} 2020, arXiv e-prints,
  arXiv:2011.08043

\bibitem[{{Xu} {et~al.}(2018){Xu}, {Wu}, {Chanussot}, {Dalla Mura}, {Bertozzi},
  \& {Wei}}]{Xu:2018}
{Xu}, Y., {Wu}, Z., {Chanussot}, J., {et~al.} 2018, IEEE Transactions on
  Geoscience and Remote Sensing, 56, 1680

\bibitem[{Yang {et~al.}(2019)Yang, Zhang, Song, \& Liu}]{Yang:2019}
Yang, Y., Zhang, J., Song, S., \& Liu, D. 2019, Remote Sensing, 11, 192

\bibitem[{Zhang {et~al.}(2018)Zhang, Li, Zhang, Chen, Feng, Jiao, \&
  Zhou}]{Zhang:2018}
Zhang, X., Li, C., Zhang, J., {et~al.} 2018, Remote Sensing, 10, 339

\bibitem[{{Zieleniewski} {et~al.}(2015){Zieleniewski}, {Thatte}, {Kendrew},
  {Houghton}, {Swinbank}, {Tecza}, {Clarke}, \& {Fusco}}]{Zieleniewski:2015}
{Zieleniewski}, S., {Thatte}, N., {Kendrew}, S., {et~al.} 2015, \mnras, 453,
  3754

\end{thebibliography}

%-------------------------------------------------------------------
\begin{appendix}
\section{Simulations of ELT/HARMONI synthetic data}
\label{ap:sim}
\begin{figure}
    \centering
    \includegraphics[width=0.5\textwidth]{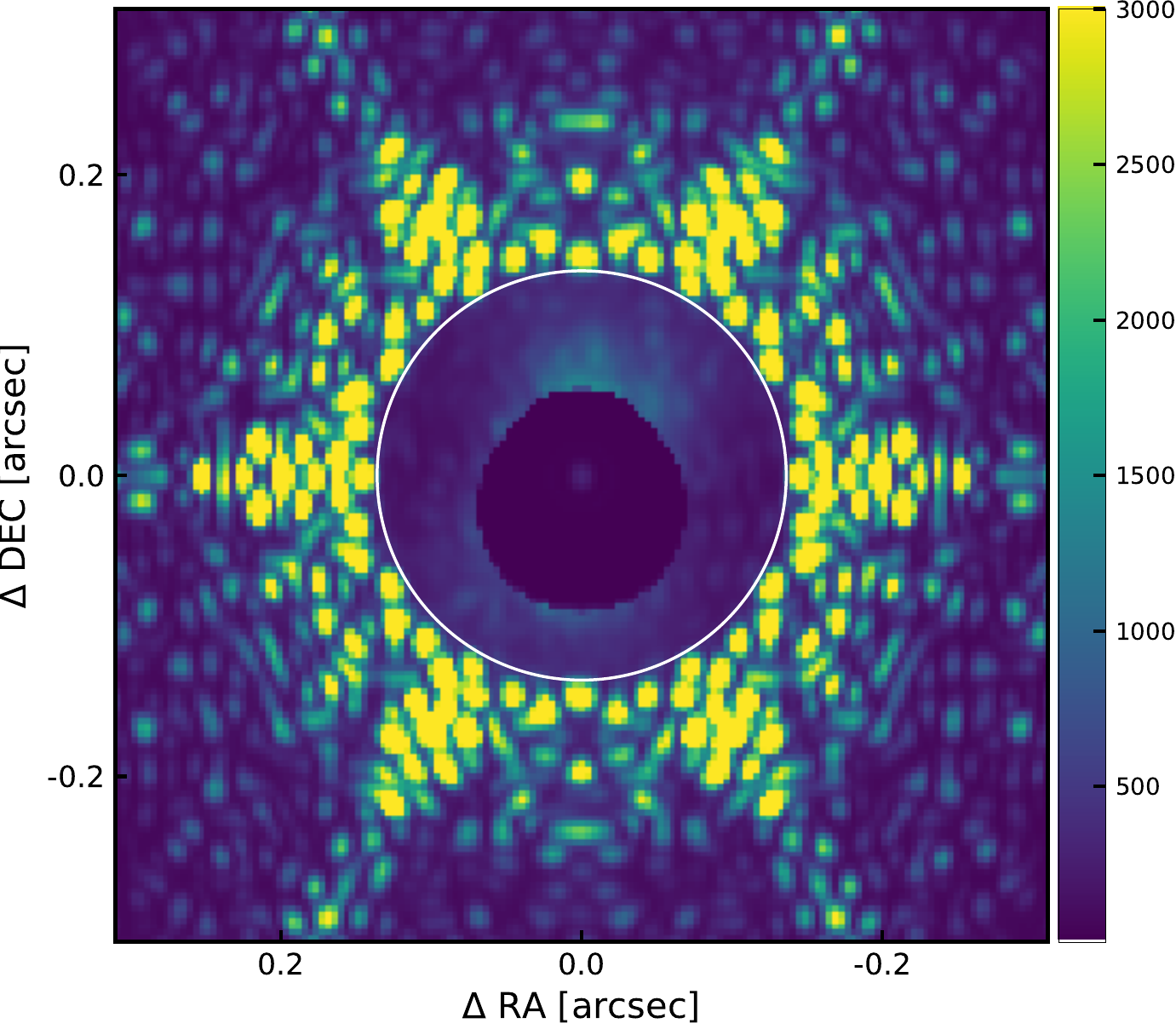}
    \caption{Simulated coronagraphic ELT/HARMONI image at $2.0\,\micron$ for an exposure of 60s with the SP1 apodizer. The central star is attenuated by the focal plane mask, which is asymmetric to take the atmospheric dispersion into account. The remaining field of view is plagued by diffracted starlight. The white line marks the adaptive optics corrected area, the region where the speckle intensity is minimized. The intensity scale is linear.}
    \label{fig:harmoni_sim}
\end{figure}

In this appendix, we briefly summarize the main characteristics of the simulation of the synthetic hyperspectral data with ELT/HARMONI. The single conjugate adaptive optics subsystem provides a Strehl ratio of 75-80\% at K band, leaving aside residuals that also take the primary mirror aberrations into account (island effect, wind shake). Additional aberrations come from the elements in the optical train toward the focal plane mask, including chromatic beam shift (due to the absence of atmospheric dispersion correction in HARMONI) and rotation of the optics in the relay system (HARMONI is mounted on the Nasmith platform). The high-contrast module yields 80 nm RMS residual aberrations in this configuration on average. The simulator produces coronagraphic PSFs for each exposure and wavelength, which are assembled in 240 data cubes of $215\times215\times1665$ pixels for a total volume of 70Gb. An off-axis PSF is also generated.

Astrophysical signal, throughput, background, and random and systematic noise were incorporated into the optical simulations to create  mock-observed data cubes, following the \texttt{HSIM} pipeline
\citep{Zieleniewski:2015}. An input scene was modeled with a central point-like young A6 star with a K=6 magnitude, with a BT-Nextgen spectrum at 8000\,K and a logarithm of the surface gravity of $\log g=4.0$ dex
\citep{Allard:2012}. The spectrum was convolved with a Gaussian line-spread function to match the resolution of the HARMONI data and was flux calibrated by the end-to-end throughput. The ESO Skycal sky model\footnote{\url{http://www.eso.org/observing/etc/bin/gen/form?INS.MODE=swspectr+INS.NAME=SKYCALC}} was used to compute the transmission of the terrestrial atmosphere at the observed airmass, further multiplied by the transmission of the ELT mirrors, the optical relay system, the instrument optics and the K-band grating, and the detector quantum efficiency\footnote{\label{foot:hsim}All curves and constants are given in the \texttt{HSIM} source code \url{https://github.com/HARMONI-ELT/HSIM}}. The input scene was convolved with the coronagraphic PSFs for each wavelength and each exposure. Background noise was further added to the data cubes, including the sky (from the same ESO Skycal sky models), the telescope wa modeled as a gray body whereby the thermal emission at the site temperature ($280.5$\,K) was multiplied with a constant emissivity\footref{foot:hsim}, the relay system and the instrument both emitting as pure blackbodies ($260.5$ and $130$\,K, respectively). Finally, Poisson noise from the observed star and the background was added along with the detector dark current ($0.0053\,\mathrm{e}^-/\mathrm{s}$), also subject to photon noise, and the read-out noise ($12\,\mathrm{e}^-$). Detector crosstalk ($2\%$) with the four contiguous pixels to each pixel was also simulated in both spatial and spectral directions, as well as broadband diffusion ($0.5\%$). Figure \ref{fig:harmoni_sim} shows a simulated exposure at $2\micron$, with the attenuated central star and the diffracted starlight, whose amplitude is minimized in the central region that  is corrected by the adaptive optics.

 \section{Distributions in the detection maps}
\label{appendix}

\begin{figure}[htbp]
    \centering
    \includegraphics[width=0.4\textwidth]{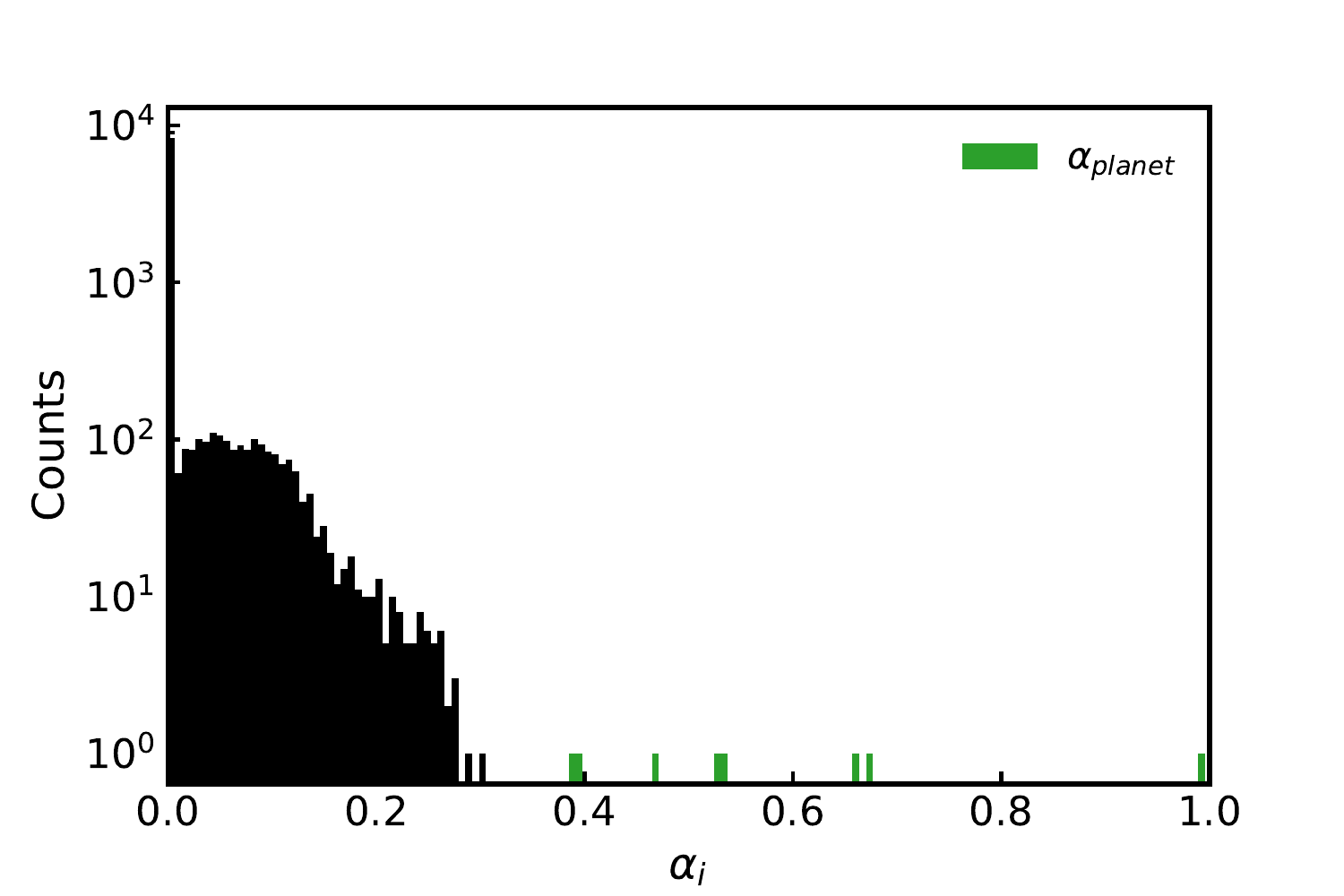}
        \includegraphics[width=0.4\textwidth]{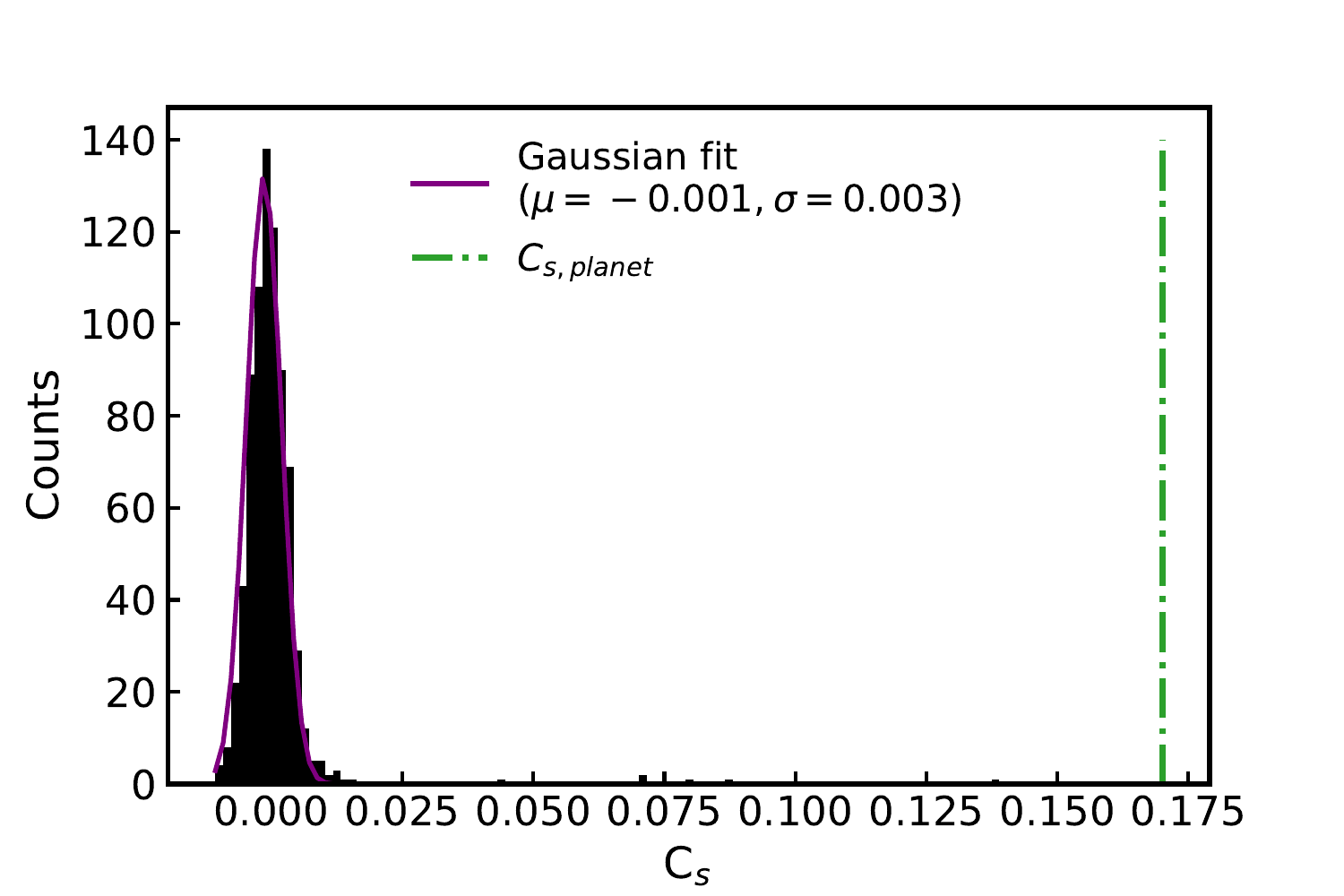}
    \caption{Distributions of spatial weights  (top) and cross-correlation strengths (bottom) after processing synthetic HARMONI data with a simulated planet at a contrast of $5\times10^{-6}$. The values at the location of the planet are highlighted in green in each histogram. The peak is enhanced for the cross-correlation for clarity. The parameters of a Gaussian fit to the distribution of the cross-correlation strengths are given.}
    \label{fig:hist_SINFONI}
\end{figure}
\begin{figure}
    \centering
    \includegraphics[width=0.4\textwidth]{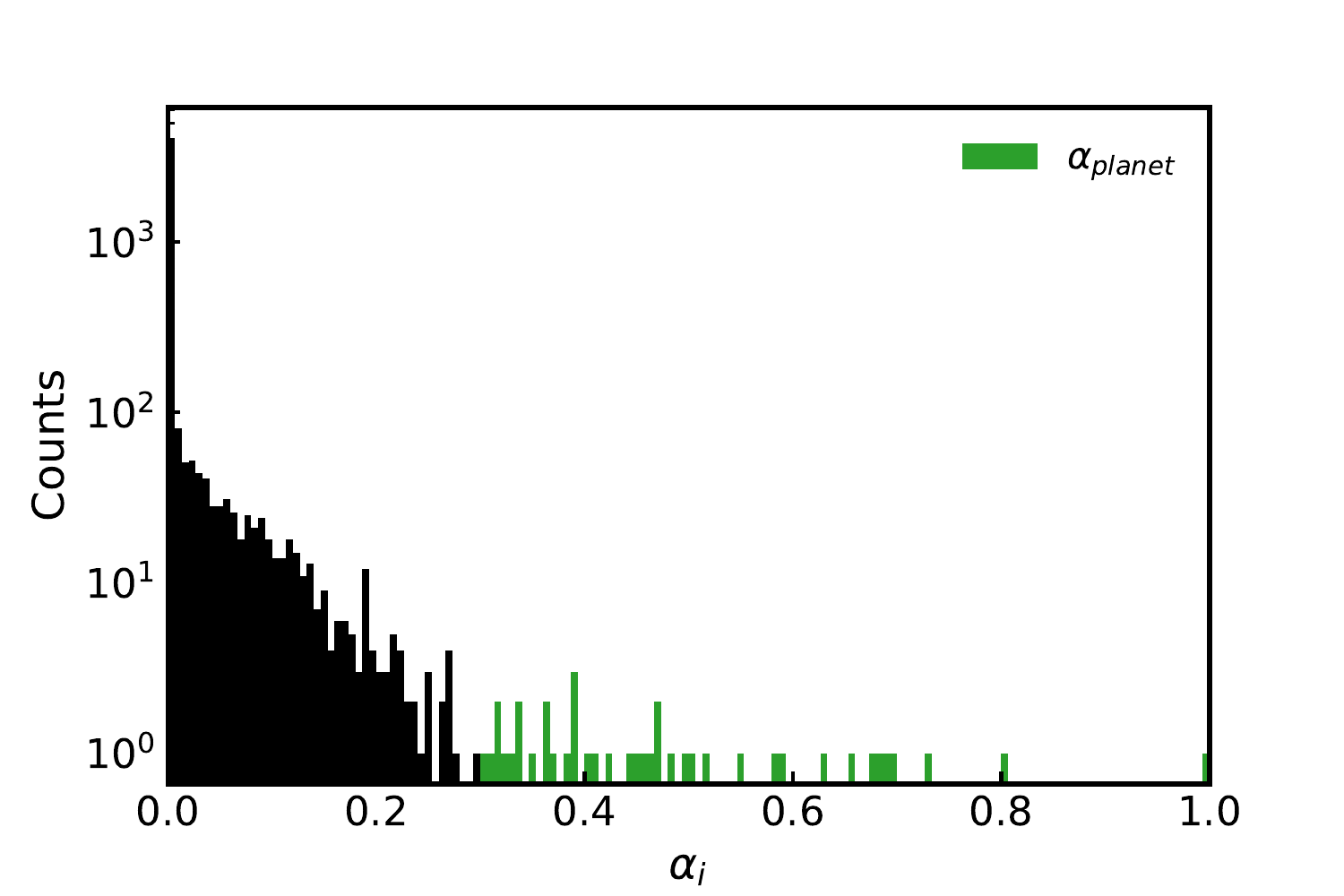}
        \includegraphics[width=0.4\textwidth]{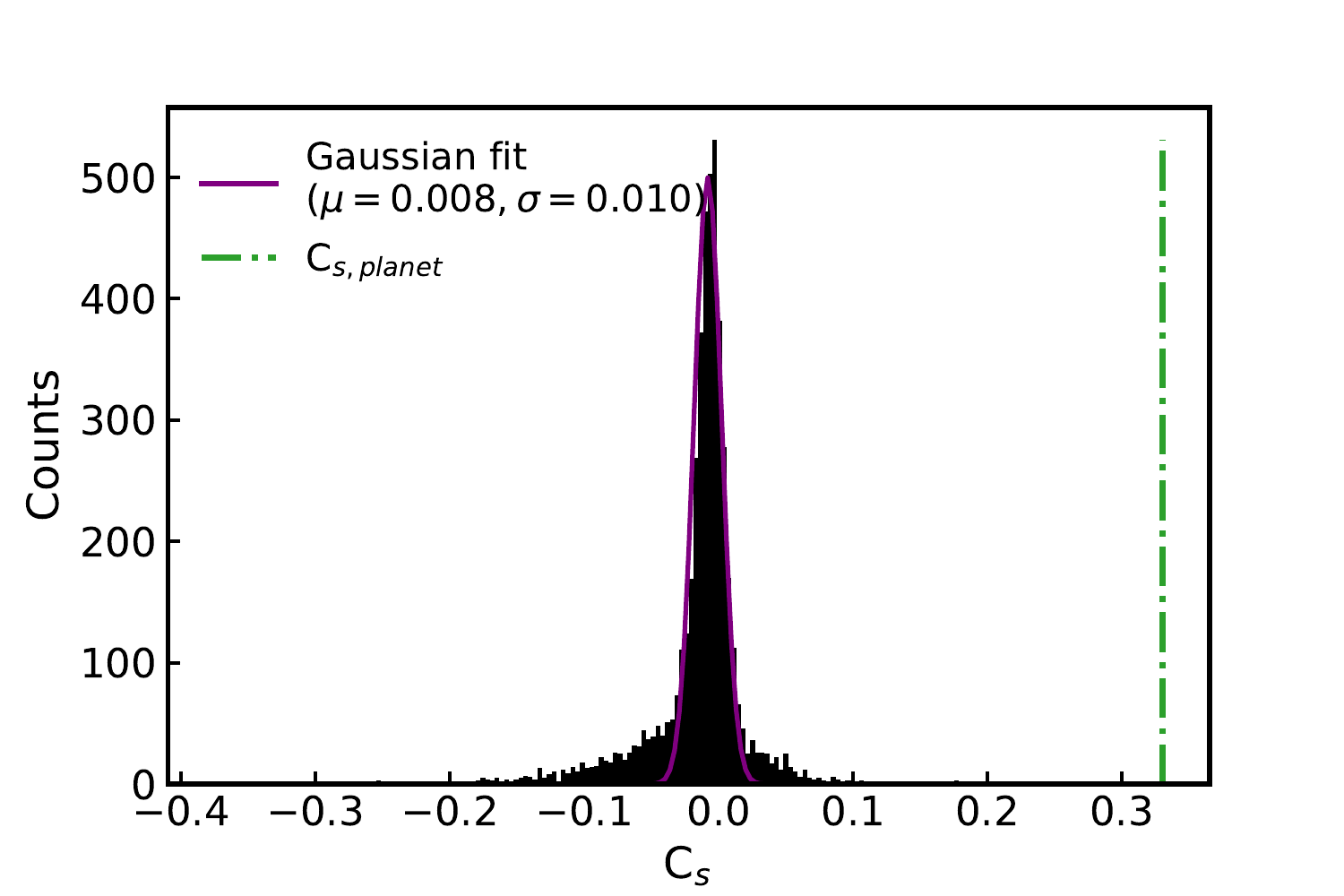}
    \caption{Same as Figure \ref{fig:hist_HARMONI}, but for the SINFONI data of $\beta$ Pictoris.}
    \label{fig:hist_HARMONI}
\end{figure}

To further illustrate the significance of the detection of the planets with spectral unmixing and cross-correlation, we built the distributions of the values in their respective maps. Figure \ref{fig:hist_HARMONI} shows the histograms corresponding to the maps for the single test case on the synthetic HARMONI data, as displayed in Figure \ref{fig:harmoni_test}. Figure \ref{fig:hist_SINFONI} illustrated the case of the on-sky SINFONI data.

In the case of spectral unmixing, the spatial weights are positive by construction. A majority of the pixels in the field of view are zeroed because they are not decomposed by the planet spectrum. A significant number of pixels have low values as a result of common starlight residuals with the planet, as discussed in the plain text. A hard threshold around $0.3$ marks the beginning of all pixels within the core of the planet, with weights up to 1. This holds for the synthetic and on-sky cases. Because of the shape of the distribution, the computed S/N cannot be converted into a false-alarm probability. \\
The cross-correlation histograms are best reproduced with a normal distribution centered on zero with very small width, although the distribution for the SINFONI shows large tails that arise from the bright structures in the map around the planet (see Figure \ref{fig:SINFONI_Bpic}). In both cases, the peak value at the planet location clearly stands out from the remaining pixels.

\section{Spectral distances}
\label{ap:dist}
With spectral unmixing, sample spectra are identified based on the dissimilarities between the spaxels within the field of view. Three metrics are commonly used in remote-sensing research: the orthogonal projection distance following Equation \ref{eq:distance}, the cross-correlation, and the spectral angle mapper, which is defined as
$$
\mathrm{SAM}(\mathbf{s_i},\mathbf{s_j})=\arccos{\Bigg(\frac{\langle\mathbf{s_i},\mathbf{s_j}\rangle}{\lVert\mathbf{s_i}\rVert\;\lVert\mathbf{s_j}\rVert}\Bigg)} \quad [\mathrm{rd}].
$$
The metric with the best spectral discriminatory power is the one that leads to the highest ratio between two similar and two distinct spectra \citep{Chang2000}. 

In order to assess that the orthogonal projection distance is the most appropriate for our purpose, we computed this ratio for the three metrics. We used the hyperspectral simulated data cube from Section \ref{sec:harmoni}. $\mathbf{s_j}$ was set as the spectrum corresponding to the spaxel at the peak of the planet PSF. Each metric was first computed with $\mathbf{s_i}$ being the spectrum of a spaxel within the core of the planet, targeting a very similar spectrum to $\mathbf{s_j}$. The metric was also computed 20 times with $\mathbf{s_i}$ as a random spectrum among all other spaxels, and then averaged to obtain a representative value for distinct spectra. The ratio of the two measurements was then calculated for all three metrics. We obtain 5.76 with the orthogonal projection distance, 4.82 with the cross-correlation, and 1.18 with $\mathrm{SAM}$. Therefore, the orthogonal projection distance is 18\% and nearly five times more effective than cross-correlation and $\mathrm{SAM}$, respectively, to distinguish between any two spectral signatures relative to the planet spectrum.

\end{appendix}

\end{document}